\documentclass[aps,prb,reprint,amsmath,amssymb,superscriptaddress]{revtex4-1}
\makeatletter
\let\frontmatter@footnote@produce\frontmatter@footnote@produce@endnote
\makeatother
\usepackage{graphicx}% Include figure files
\usepackage{physics}
\usepackage{bm}  %\vec{} for bold font in math
\usepackage{amsmath} %\text{} for text in math
\usepackage{braket} %\braket{ bra | V | ket } 
\usepackage[colorlinks,allcolors=blue,urlcolor=blue]{hyperref}% add hypertext capabilities

\bibpunct{[}{]}{,}{n}{}{}

\newcommand{\rr}{\mathbf{r}}

\newcommand{\kk}{\bm{k}}
\renewcommand{\qq}{\bm{q}}

\newcommand{\ttau}{\bm{\tau}}

\renewcommand{\vec}[1]{\mathbf{#1}}

\begin{document}
\title{Long-range quadrupole electron-phonon interaction from first principles}
\author{Jinsoo Park}
\author{Jin-Jian Zhou}
\author{Vatsal A. Jhalani}
\affiliation{Department of Applied Physics and Materials Science, California Institute of Technology, Pasadena, California 91125, USA.}
\author{Cyrus E. Dreyer}
\affiliation{Department of Physics and Astronomy, Stony Brook University, Stony Brook, New York 11794-3800}
\affiliation{Center for Computational Quantum Physics, Flatiron Institute, 162 Fifth Avenue, New York, New York 10010}
\author{Marco Bernardi}
\email[Corresponding author: ]{bmarco@caltech.edu}
\affiliation{Department of Applied Physics and Materials Science, California Institute of Technology, Pasadena, California 91125, USA.}
%\date{\today}
%
% ABSTRACT
%
\begin{abstract}
Lattice vibrations in materials induce perturbations on the electron dynamics in the form of long-range (dipole and quadrupole) and short-range (octopole and higher) potentials. 
The dipole Fr\"ohlich term can be included in current first-principles electron-phonon ($e$-ph) calculations and is present only in polar materials.  
The quadrupole $e$-ph interaction is present in both polar and nonpolar materials, but currently it cannot be computed from first principles.
Here we show an approach to compute the quadrupole $e$-ph interaction and include it in \textit{ab initio} calculations of $e$-ph matrix elements. 
The accuracy of the approach is demonstrated by comparing with direct density functional perturbation theory calculations. 
We apply our method to silicon as a case of a nonpolar semiconductor and tetragonal PbTiO$_3$ as a case of a polar piezoelectric material. 
In both materials we find that the quadrupole term strongly impacts the $e$-ph matrix elements. 
Analysis of $e$-ph interactions for different phonon modes reveals that the quadrupole term mainly affects optical modes in silicon and acoustic modes in PbTiO$_3$, 
although  the quadrupole term is needed for all modes to achieve quantitative accuracy. 
The effect of the quadrupole $e$-ph interaction on electron scattering processes and transport is shown to be important. 
Our approach enables accurate studies of $e$-ph interactions in broad classes of nonpolar, polar and piezoelectric materials.
\end{abstract} 
\maketitle
%
% ---------------- INTRODUCTION  ------------------------
%
\section{Introduction}
\vspace{-10pt}
Electron-phonon ($e$-ph) interactions are key to understanding electrical transport, nonequilibrium dynamics, and superconductivity~\cite{zimanElectrons2001}. 
First-principles calculations can provide microscopic insight into $e$-ph scattering processes and are rapidly emerging as a quantitative tool for investigating charge transport and ultrafast carrier dynamics in materials~\cite{Bernardi2016, bernardiInitio2014, Zhou2016, jhalaniUltrafast2017, leeCharge2018, Zhou2018, zhouPredicting2019, Perturbo, liElectrical2015, liuFirstprinciples2017, maFirstprinciples2018, sohierMobility2018}. 
The typical workflow combines density functional theory (DFT)~\cite{martinElectronic2004} calculations of the ground state and band structure 
with density functional perturbation theory (DFPT)~\cite{Baroni-DFPT} for phonon dispersions and $e$-ph perturbation potentials. 
As DFPT can compute the electronic response to periodic lattice perturbations (phonons) with arbitrary wave-vector $\qq$, 
the DFPT framework can capture both short- and long-range $e$-ph interactions.
\\
\indent  
However, a key challenge is that DFPT is too computationally demanding to be carried out on the fine Brillouin zone grids needed to compute electron scattering rates and transport properties. 
The established approach in first-principles $e$-ph studies~\cite{Perturbo} is to carry out DFPT calculations on coarse Brillouin zone grids with of order 10$\times$10$\times$10 $\qq$-points, 
followed by interpolation of the $e$-ph matrix elements with a localized basis set such as Wannier functions or atomic orbitals~\cite{Agapito2018}. 
As the perturbation potential can be non-analytic near $\qq \!=\! 0$~\cite{Agapito2018} or even exhibit a divergence for certain phonon modes, 
interpolation is particularly challenging and less reliable in the region between $\qq=0$ and its nearest-neighbor $\qq$-points in the coarse DFPT grid.  
This small-$\qq$ region is critical as it is dominated by long-range $e$-ph interactions, whose treatment can affect the quality of the interpolation even at larger values of $q$ in the Brillouin zone. 
\\ 
\indent
A multipole expansion of the $e$-ph perturbation potential shows that in the long-wavelenght limit (phonon wave-vector $\qq \!\rightarrow\! 0$)   
the long-range dipole Fr\"ohlich term diverges as $1/q$, the quadrupole term approaches a constant value and the short-range octopole and higher terms vanish~\cite{Lawaetz,Vogl1976}. 
These trends in momentum space are due to the spatial decay of the $e$-ph interactions, with $1/r^2$ trend for the dipole, $1/r^3$ for the quadrupole, and $1/r^4$ or faster for the short-range part. 
The open question is how one can carry out the $e$-ph matrix element interpolation in the region near $\qq\!=\!0$ using analytical expressions for the long-range dipole and quadrupole terms. 
These expressions have been obtained by Vogl~\cite{Vogl1976}, but need to be rewritten in the \textit{ab initio} formalism and computed with first-principles quantities such as 
the atomic dynamical dipoles~\cite{Baroni-DFPT} and quadrupoles~\cite{Stengel2013, Dreyer2018, Royo2019} induced by lattice vibrations, which can be computed with DFPT. 
For each atom $\kappa$, one can obtain Born charge ($\mathbf{Z}_{\kappa}$) and dynamical quadrupole ($\vec{Q}_{\kappa}$) tensors, which, once contracted with
the phonon eigenvector, give the atomic contributions to the dipole and quadrupole $e$-ph interactions. 
\\
\indent
The dipole Fr\"ohlich term has been derived following this strategy~\cite{sjaksteWannier2015,verdiFr2015} and employed in electron scattering rate and transport calculations~\cite{Zhou2016,Zhou2018}.  
The quadrupole term has not yet been derived or implemented in first-principles calculations and its important effect on the $e$-ph matrix elements has been overlooked. 
To understand the role of long-range dipole and quadrupole $e$-ph interactions, it is useful to consider separately the interactions for different phonon modes in the long-wavelength limit, 
discerning the effect of longitudinal and transverse, and acoustic and optical modes. Analytical models of $e$-ph interactions rely on such an intuition for the role of different phonon modes in various materials~\cite{mahanCondensed2011}. 
\\
\indent
In ionic and polar covalent crystals (here and below, denoted as polar materials), the dipole Fr\"ohlich term is dominant as $\qq \rightarrow 0$ due to its $1/q$ trend. 
This $e$-ph interaction is due to longitudinal optical (LO) phonons and it dominates small-$\qq$ scattering. For other phonon modes in polar materials, the dipole term vanishes,  
and the dominant long-range $e$-ph interaction is the quadrupole term, which is particularly important for acoustic phonons in piezoelectric (polar noncentrosymmetric) materials. 
In nonpolar semiconductors such as silicon and germanium, the dipole Fr\"ohlich interaction vanishes and the quadrupole term is a key contribution for all modes in the long-wavelength limit. %In centrosymmetric nonpolar crystals, 
The quadrupole $e$-ph interaction is thus expected to play an important role in many classes of materials, making a compelling case for its inclusion in the first-principles framework.
\\
\indent
%
% HERE WE SHOW
% 
Here we show \emph{ab initio} calculations of the long-range quadrupole $e$-ph interaction and an approach to include it in the $e$-ph matrix elements. 
The accuracy of our method is confirmed by comparing the $e$-ph matrix elements with direct DFPT calculations. 
We find that the quadrupole contribution is significant for most phonon modes in both nonpolar and polar materials. 
In silicon, a nonpolar semiconductor, the quadrupole term has a large effect on the $e$-ph coupling for optical modes, but is negligible for acoustic modes in the long-wavelength limit.   
In tetragonal $\text{PbTiO}_3$, a polar piezoelectric material, the quadrupole corrections are substantial for all phonon modes and particularly important for acoustic modes, which contribute to the piezoelectric $e$-ph interaction. 
Including only the long-range Fr\"ohlich interaction and neglecting the quadrupole term leads to large errors in $\text{PbTiO}_3$, while adding the quadrupole term leads to $e$-ph matrix elements that accurately reproduce the DFPT benchmark results for all phonon modes in the entire Brillouin zone.
We investigate the impact of the quadrupole $e$-ph interaction on the electron scattering rates and mobility in silicon and $\text{PbTiO}_3$, finding mobility corrections of order 10\% in silicon and 20\% in $\text{PbTiO}_3$ at 100~K 
(and smaller corrections at 300~K) when the quadrupole term is included. The correction on the scattering rate at low electron energy in $\text{PbTiO}_3$ is substantial. 
Taken together, our results highlight the need to include the quadrupole term in all materials to correctly capture the long-range $e$-ph interactions.  
In turn, this development enables more precise calculations of electron dynamics and scattering processes from first principles. 
\begin{figure}[!h]
\includegraphics[width=1.0\columnwidth]{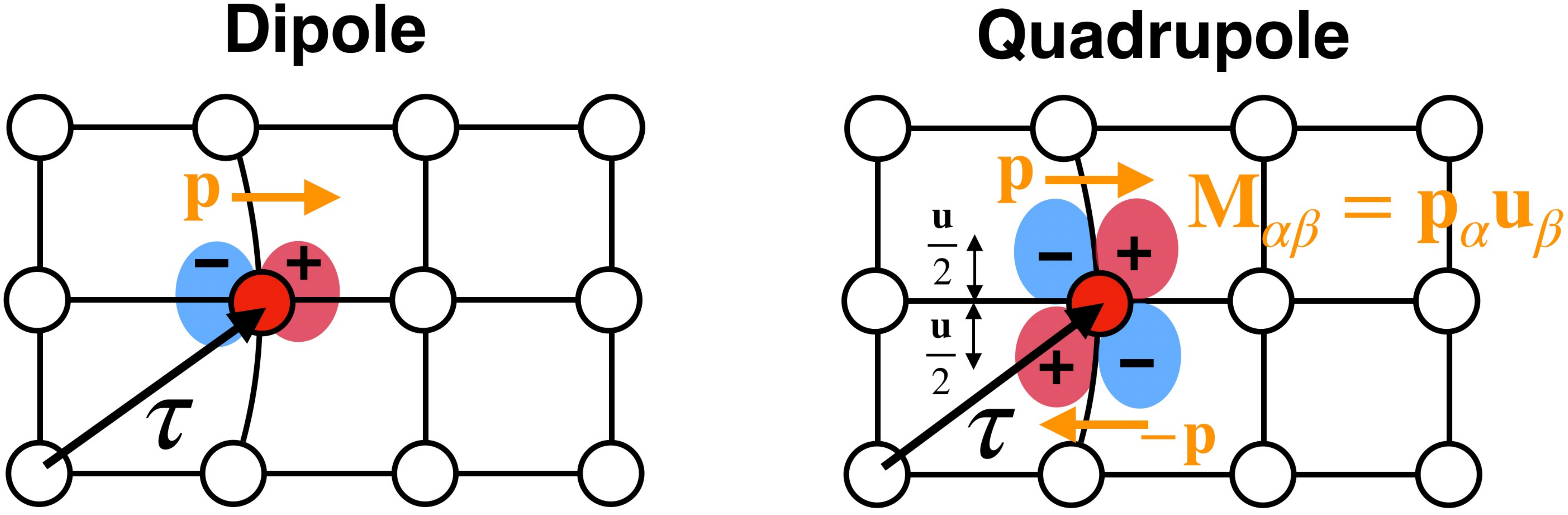}
\caption{Schematic of the dipole and quadrupole charge configurations giving rise to long-range $e$-ph interactions.
}\label{fig:schematic}
\end{figure}
%
%%%  ------------------  SECTION: THEORY  ------------------------
%
\section{Theory}
The electron distribution changes in response to a displacement of an atom from its equilibrium position.  
The cell-integrated charge response to a displacement of atom $\kappa$ due to a phonon with wave-vector $\qq \rightarrow 0$ can be written as a multipole expansion~\cite{Royo2019}:
% CHARGE RESPONSE
\begin{equation}\label{eq:charge_multipole_expansion}
C^{\qq}_{\kappa,\alpha} = -i Z_{\kappa,\alpha\beta}\, q_\beta - \frac{1}{2} Q_{\kappa,\alpha\beta\gamma}\,q_\beta q_\gamma  + \dots,
\end{equation}
where summation over the Cartesian indices $\beta$ and $\gamma$ is implied. 
% TENSORS
This polarization response defines the Born effective charge $\bf{Z}_{\kappa}$, a rank-2 tensor associated with the dipole term, and the dynamical quadrupole $\mathbf{Q}_{\kappa}$, the rank-3 tensor in the quadrupole term;    
both tensors can be computed in the DFPT framework~\cite{Baroni-DFPT, Stengel2013}. 
Each of the dipole and quadrupole responses generates macroscopic electric fields and corresponding long-range $e$-ph interactions in semiconductors and insulators~\cite{frohlichElectrons1954, Vogl1976}, 
while in metals they are effectively screened out. 
\\
\indent
In a field-theoretic treatment of the $e$-ph interactions, one computes the dipole and quadrupole perturbation potentials $\Delta V_{\nu \qq}$ due to a phonon with mode index $\nu$ and wave-vector $\qq$, and the corresponding 
$e$-ph matrix elements~\cite{Bernardi2016}
% MATRIX ELEMENT (GENERAL) 
\begin{equation} \label{eq:me}
g_{mn\nu}(\kk,\qq) = \left(\frac{\hbar}{2\omega_{\nu \qq}}\right)^{\frac{1}{2}} \bra{m\kk+\qq} \Delta V_{\nu \qq} \ket{n\kk}, 
\end{equation}
which quantify the probability amplitude of an electron in a Bloch state $\ket{n\kk}$ with band index $n$ and crystal momentum $\kk$ to scatter into a final state $\ket{m\kk+\qq}$ 
by emitting or absorbing a phonon with energy $\hbar \omega_{\nu \qq}$.
%
% SUBSECTION: DIPOLE / QUADRUPOLE DERIVATIONS
%
\subsection{Dipole and quadrupole $e$-ph interactions}
\indent
To derive the dipole and quadrupole perturbation potentials, we consider a Born-von Karman (BvK) crystal~\cite{bornDynamical1954} with $N$ unit cells and volume $N\Omega$. 
The potential due to a dipole configuration with dipole moment $\vec{p}$ centered at position $\ttau$ in the crystal (see Fig.~\ref{fig:schematic}) can be written as~\cite{sjaksteWannier2015,verdiFr2015}
% DIPOLE POTENTIAL 
\begin{equation}\label{eq:pot_dip}
\Delta V^\text{dip}(\vec{r};\ttau) = i\frac{e}{N\Omega\, \varepsilon_{0}} \sum_{\qq} \sum_{\vec{G}\neq-\qq}\frac{\vec{p}\cdot(\qq+\vec{G})e^{i(\qq+\vec{G})\cdot(\vec{r}-\ttau)}}{(\qq+\vec{G})\cdot\boldsymbol{\epsilon}\cdot(\qq+\vec{G})},
\end{equation}
where $\boldsymbol{\epsilon}$ is the dielectric tensor of the material, the phonon wave-vector $\qq$ belongs to a regular Brillouin zone grid with $N$ points, 
and $\vec{G}$ are reciprocal lattice vectors. 
This result is derived by adding together the potentials generated in the crystal by two point charges of opposite sign with distance $\vec{u} \rightarrow 0$,  
resulting in a dipole $\vec{p}$~\cite{verdiFr2015,Zhou2016}. 
\\
\indent 
% ATOMIC DIPOLE POTENTIAL 
The potential in Eq.~(\ref{eq:pot_dip}) is readily extended to the case of an atomic dynamical dipole $\vec{p}_{\kappa,\vec{R}}$ from atom $\kappa$ in the unit cell at Bravais lattice vector $\vec{R}$, 
due to the displacement induced by a phonon with mode index $\nu$ and wave-vector $\qq$. 
The resulting atomic dynamical dipole is $\vec{p}_{\kappa,\vec{R}} = (e\vec{Z}_\kappa)\, \tilde{\vec{e}}^{(\kappa)}_{\nu \qq} e^{i \vec{q} \cdot \vec{R}}$, where the phonon eigenvector projected on atom $\kappa$ is defined as  
$\tilde{\vec{e}}^{(\kappa)}_{\nu \qq} = \vec{e}^{(\kappa)}_{\nu \qq} / \sqrt{M_\kappa}$, with $\vec{e}_{\nu \qq}$ the eigenvector of the dynamical matrix at $\qq$ and $M_{\kappa}$ the mass of atom $\kappa$.  
Summing over the contributions from all atoms $\kappa$ at lattice vectors $\mathbf{R}$ with positions $\ttau_{\kappa \vec{R}}= \ttau_{\kappa} + \vec{R}$ in the BvK supercell,  
the total $e$-ph dipole interaction due to the phonon mode is $\Delta V^\text{dip}_{\nu \qq} (\rr)  = \sum_{\kappa \vec{R}} \Delta V^\text{dip} (\vec{r};\ttau_{\kappa \vec{R}})$. 
Using the identity $\frac{1}{N} \sum_\vec{R} e^{i \vec{q} \cdot \vec{R} } = \delta_{\vec{q}, 0}$, we obtain: 
\begin{equation}\label{eq:atomic-dipole}
\Delta V^\text{dip}_{\nu \vec{q}}(\vec{r}) = i\frac{e^2}{\Omega \varepsilon_{0}}\! \sum_\kappa M_\kappa^{-1/2} \! \sum_{\vec{G}\neq-\qq} \frac{(\vec{Z_\kappa} \vec{e}^{(\kappa)}_{\nu \qq}) \!\cdot\! (\qq+\vec{G}) e^{i(\qq+\vec{G})\cdot(\vec{r}-\ttau)}}{(\qq+\vec{G})\cdot\boldsymbol{\epsilon}\cdot(\qq+\vec{G})}
\end{equation}
% 
% DIPOLE MATRIX ELEMENT  
The \textit{ab initio} Fr\"ohlich $e$-ph coupling is obtained by evaluating the matrix elements with this potential: 
\begin{widetext}
\begin{equation}\label{eq:gdip}
g^{\rm dip}_{mn\nu}(\kk,\qq) = i\frac{e^2}{\Omega \varepsilon_{0}}\! \sum_\kappa \left( \frac{\hbar}{2\omega_{\nu \qq} M_\kappa}  \right)^{1/2} 
\sum_{\vec{G}\neq-\qq} \frac{(\vec{Z_\kappa} \vec{e}^{(\kappa)}_{\nu \qq}) \!\cdot\! (\qq+\vec{G})} {(\qq+\vec{G})\cdot\boldsymbol{\epsilon}\cdot(\qq+\vec{G})} 
\bra{m\kk+\qq} e^{i(\qq+\vec{G})\cdot(\vec{r}-\ttau)} \ket{n\kk}.
\end{equation}
\end{widetext}

\indent
% QUADRUPOLE POTENTIAL
The potential due to the dynamical quadrupole response can be derived with a similar strategy. 
We first consider the potential generated by a quadrupole charge configuration consisting of two equal and oppositely oriented dipoles  
$\vec{p}$ and $-\vec{p}$, centered at positions $\ttau \pm \frac{\vec{u}}{2}$ respectively (see Fig.~\ref{fig:schematic}). 
The configuration, with quadrupole moment~\cite{Griffiths} $M_{\alpha \beta} = p_\alpha u_\beta$, gives a potential: 
\vspace{-9pt}
 \begin{equation}\label{eq:pot_quad_full}
 \begin{split}
\Delta V^\text{quad}(\vec{r};\ttau) \!=\! \lim_{\vec{u}\rightarrow0}&\! \left[ \Delta V^\text{dip}\left(\vec{r};\ttau+\frac{\vec{u}}{2}\right) \!-\! \Delta V^\text{dip}\left(\vec{r};\ttau-\frac{\vec{u}}{2}\right)\! \right]\\
= \frac{e}{N\Omega\, \varepsilon_{0}} \sum_{\qq} \sum_{\vec{G}\neq-\qq}& \frac{ (\qq+\vec{G}) \cdot \vec{M} \cdot (\qq+\vec{G}) }{ (\qq+\vec{G})\cdot\boldsymbol{\epsilon}\cdot(\qq+\vec{G}) }\,e^{i(\qq+\vec{G})\cdot(\vec{r}-\ttau)},
\end{split}
 \end{equation}
where to obtain the second line we used $\Delta V^\text{dip}(\vec{r};\ttau)$ in Eq.~(\ref{eq:pot_dip}) and expanded the first line to first order in $\vec{u}$.\\
\indent
% ATOMIC QUADRUPOLE POTENTIAL 
Similar to the dipole case, the potential from atomic quadrupoles $(\vec{M}_{\kappa,\vec{R}})_{\alpha \beta} \!=\! \frac{1}{2}(e\vec{Q}_{\kappa})_{\alpha \beta \gamma}\, \vec{e}^{(\kappa)}_{\nu \qq,\gamma}\, e^{i \vec{q} \cdot \vec{R}}$ 
due to the displacement induced by a phonon is obtained as $\Delta V^{\rm quad}_{\nu \qq} (\rr)  = \sum_{\kappa \vec{R}} \Delta V^\text{quad} (\vec{r};\ttau_{\kappa \vec{R}})$. 
Following steps analogous to the dipole case, we find:  
\begin{equation}
\begin{split}
\Delta V^\text{quad}_{\nu\qq} (\rr)  = \frac{e^2}{\Omega\varepsilon_{0}} \sum_{\kappa} M_\kappa^{-1/2} \!\sum_{\vec{G}\neq-\qq} &\frac{1}{2} \frac{(\qq + \vec{G}) \cdot ( \vec{Q}_{\kappa} \vec{e}_{\nu \qq}^{(\kappa)} ) \cdot (\qq + \vec{G})}
{(\qq+\vec{G})\cdot\boldsymbol{\epsilon}\cdot(\qq+\vec{G})} \\
& \times e^{ i(\qq+\vec{G})\cdot(\vec{r}-\ttau_{\kappa}) }.
\end{split}
\end{equation}

% QUADRUPOLE MATRIX ELEMENT
The corresponding $e$-ph matrix elements due to the quadrupole perturbation potential are: 
\begin{widetext}
\begin{equation}\label{eq:gquad}
g_{mn\nu}^\text{quad}(\kk,\qq) = \frac{e^2}{\Omega \varepsilon_{0}} \sum_{\kappa} \left( \frac{\hbar}{2\omega_{\nu \qq} M_\kappa} \right)^{\frac{1}{2}} \sum_{\vec{G}\neq-\qq} \frac{1}{2} 
\frac{ (\qq + \bm{G})_\alpha (Q_{\kappa,\alpha\beta\gamma} 
\mathrm{e}_{\nu \qq, \gamma}^{(\kappa) } ) (\qq + \bm{G})_\beta }{ (\qq + \bm{G})_\alpha \epsilon_{\alpha\beta} (\qq + \bm{G})_\beta}
\bra{m\kk+\qq} e^{i(\qq+\vec{G})\cdot(\vec{r}-\ttau_\kappa)} \ket{n\kk}.
\end{equation}
\end{widetext}
Note that in the $\qq \!\rightarrow\! 0$ limit the Fr\"ohlich $e$-ph matrix elements are of order $1/q$ and the quadrupole matrix elements of order $q^0$, thus approaching a constant value; both quantities are non-analytic as $\qq \rightarrow 0$. 
Octopole and higher electronic responses in Eq.~(\ref{eq:charge_multipole_expansion}) lead to potentials that vanish as $\qq \!\rightarrow\! 0$ and can be grouped together into a short-range $e$-ph interaction, 
commonly referred to as the \lq\lq deformation potential'' in analytic $e$-ph theories~\cite{Vogl1976}.
% 
% SUBSECTION: INTERPOLATION SCHEME
%
\subsection{Interpolation scheme for $e$-ph interactions}
\indent 
% TOTAL E-PH MATRIX ELEMENT
The total $e$-ph matrix elements $g$ (here we omit the band and mode indices) can be formed by adding together the short-range part $g^\text{S}$ and the dipole and quadrupole interactions, 
which can be combined into a long-range part $g^\text{L}$. 
Therefore, 
\begin{equation}\label{eq:gtot_gs_gl}
\begin{split}
g =&\,\, g^\text{S} + g^\text{L} \\
             =&\,\, g^\text{S} + g^\text{dip} + g^\text{quad}.
\end{split}
\end{equation}

We start from a set of $e$-ph matrix elements $g(\kk,\qq)$ computed with DFPT on a regular coarse grid of $\kk$- and $\qq$-points~\cite{Baroni-DFPT}. 
The short-range part is obtained by subtracting the long-range terms on the coarse grid, $g^\text{S}(\kk,\qq) = g(\kk,\qq) - g^\text{dip}(\kk,\qq) - g^\text{quad}(\kk,\qq)$.
The short-range $e$-ph matrix elements decay rapidly in real space, and thus are ideal for interpolation using a localized basis set such as Wannier functions~\cite{pizziWannier902020} or atomic orbitals~\cite{Agapito2018}. 
After interpolating the short-ranged part~\cite{Perturbo} on fine $\kk$- and $\qq$-point grids, we add back the long-range dipole and quadrupole matrix elements, computed using Eqs.~(\ref{eq:gdip}) and (\ref{eq:gquad})   
directly at the fine-grid $\kk$ and $\qq$-points. 
\\
\indent
% GAUGE INVARIANT E-PH MATRIX ELEMENT
As DFPT accurately captures the long-range dipole and quadrupole $e$-ph interactions~\cite{Baroni-DFPT}, the matrix elements obtained from DFPT can be used as a benchmark for the interpolated results.  
For this comparison, following Ref.~\onlinecite{sjaksteWannier2015} we compute the gauge-invariant $e$-ph coupling strength, $D^\nu_\text{tot}(\qq)$, which is proportional to the absolute value of the $e$-ph matrix elements:
\begin{equation}\label{eq:deformation_pot}
D^\nu_\text{tot}(\qq)= \sqrt{\frac{2\omega_{\nu\qq}M_{\rm uc}}{\hbar^2}\sum_{mn} \frac{\abs{g_{mn\nu}(\kk=\Gamma,\qq)}^2}{N_b} },
\end{equation} 
where $M_{\rm uc}$ is the mass of the unit cell and the band indices $n$ and $m$ run over the $N_b$ bands selected for the analysis. 
%
% SUBSECTION:  COMPUTATIONAL DETAILS
%
\subsection{Computational details} 
% DFT AND DFPT
We investigate the effect of the quadrupole $e$-ph interaction in silicon, a nonpolar semiconductor, and tetragonal $\text{PbTiO}_3$, a polar piezoelectric material. Calculations on GaN are shown in the companion work~\cite{Jhalani2020}. 
The ground state and band structure are obtained using DFT in the local density approximation with a plane-wave basis using the {\sc Quantum ESPRESSO} code~\cite{giannozziQUANTUM2009}. 
Kinetic energy cutoffs of 40~Ry for silicon and 76~Ry for $\text{PbTiO}_3$ are employed, together with scalar-relativistic norm-conserving pseudopotentials from Pseudo Dojo~\cite{vansettenPseudoDojo2018}.
The calculations employ lattice constants of 10.102~bohr for silicon and 7.275~bohr (with aspect ratio $c/a=$ 1.046) for $\text{PbTiO}_3$. We use the dynamical quadrupole tensors computed in Ref.~\onlinecite{Royo2019}. 
The phonon dispersions and $e$-ph perturbation potentials on coarse $\qq$-point grids are computed with DFPT~\cite{Baroni-DFPT}.   
We employ the {\sc perturbo} code~\cite{Perturbo} to compute the $e$-ph matrix elements on coarse Brillouin zone grids with 
$10\times10\times10$ $\kk$- and $\qq$-points for silicon and $8\times8\times8$ $\kk$- and $\qq$-points for $\text{PbTiO}_3$. 
The Wannier functions are computed with the {\sc Wannier90} code~\cite{pizziWannier902020} and employed in {\sc perturbo}~\cite{Perturbo} to interpolate the short-range $e$-ph matrix elements. 
\\
\indent
% SC RATE
We compute the scattering rates and electron mobility using the {\sc perturbo} code~\cite{Perturbo}. Briefly, the band- and $\kk$- dependent $e$-ph scattering rate $\Gamma_{n\kk}$ is obtained as
\begin{equation}\label{eq:scatrate}
\begin{split}
\Gamma_{n\kk}=&\frac{2\pi}{\hbar}\sum_{m \nu \qq}\abs{g_{mn\nu}(\kk,\qq)}^2\\
&[(N_{\nu \qq}+1-f_{m\vec{k+q}})\delta(\varepsilon_{n\kk}-\varepsilon_{m\vec{k+q}}-\hbar\omega_{\nu\qq}) \\
&~+ (N_{\nu \qq}+f_{m\vec{k+q}})\delta(\varepsilon_{n\kk}-\varepsilon_{m\vec{k+q}} +\hbar\omega_{\nu\qq})],
\end{split}
\end{equation}
where  $\varepsilon_{n\kk}$ and $\hbar\omega_{\nu \qq}$ are the electron and phonon energies, respectively, and $f_{n\kk}$ and $N_{\nu \qq}$ the corresponding temperature-dependent occupations. 
The scattering rate can be further divided into the long-range part~\cite{Zhou2016}, $\Gamma_{n\kk}^{L}$, by replacing $\abs{g}^2$ in Eq.~(\ref{eq:scatrate}) with $\abs{g^\text{L}}^2$. 
% MOBILITY
The carrier mobility is computed using $\mu=\sigma/(n_c e)$, where $\sigma$ is the electrical conductivity and $n_c$ is the carrier concentration. 
The electrical conductivity $\sigma$ is computed within the relaxation time approximation of the Boltzmann transport equation~\cite{mahanManyParticle2000,Perturbo}:
\begin{equation}
\sigma_{\alpha\beta} = e^2 \int_{-\infty}^{+\infty} {dE (-\partial f / \partial E) \Sigma_{\alpha\beta}(E,T)},
\end{equation}
where $\Sigma_{\alpha\beta}(E,T)$ is the transport distribution function at energy $E$,
\begin{equation}
    \Sigma_{\alpha\beta}(E,T)=\frac{s}{ \mathcal{N}_{\kk} \Omega} \sum_{n\kk} \tau_{n\kk}(T) v_{n\kk}^\alpha v_{n\kk}^\beta \delta(E-\varepsilon_{n\kk}),
\end{equation}
which is computed in {\sc perturbo} using the tetrahedron integration method~\cite{blochlImproved1994}. Above, $s$ is the spin degeneracy, $\mathcal{N}_{\kk}$ is the number of $\kk$-points, $v_{n\kk}$ is the band velocity, and $\tau_{n\kk}=(\Gamma_{n\kk})^{-1}$ is the relaxation time. 
The mobility is computed with non-degenerate electron concentrations of $10^{15}~\text{cm}^{-3} $ for silicon and $10^{17}~\text{cm}^{-3}$ for $\text{PbTiO}_3$. 
To fully converge the scattering rates and mobility, we use $e$-ph matrix elements evaluated on fine Brillouin zone grids with $200\times200\times200$ $\kk$-points and $8\times 10^6$ random $\qq$-points. 

%
%SECTION:   --------------  RESULTS AND DISCUSSION  -----------------------------------
%
\section{Results}
\subsection{Quadrupole effect on the $e$-ph matrix elements} 
%
% -------------  FIGURE 1:  E-PH SILICON  -----------------
%
\begin{figure}[!t]
\includegraphics[width=0.95\columnwidth]{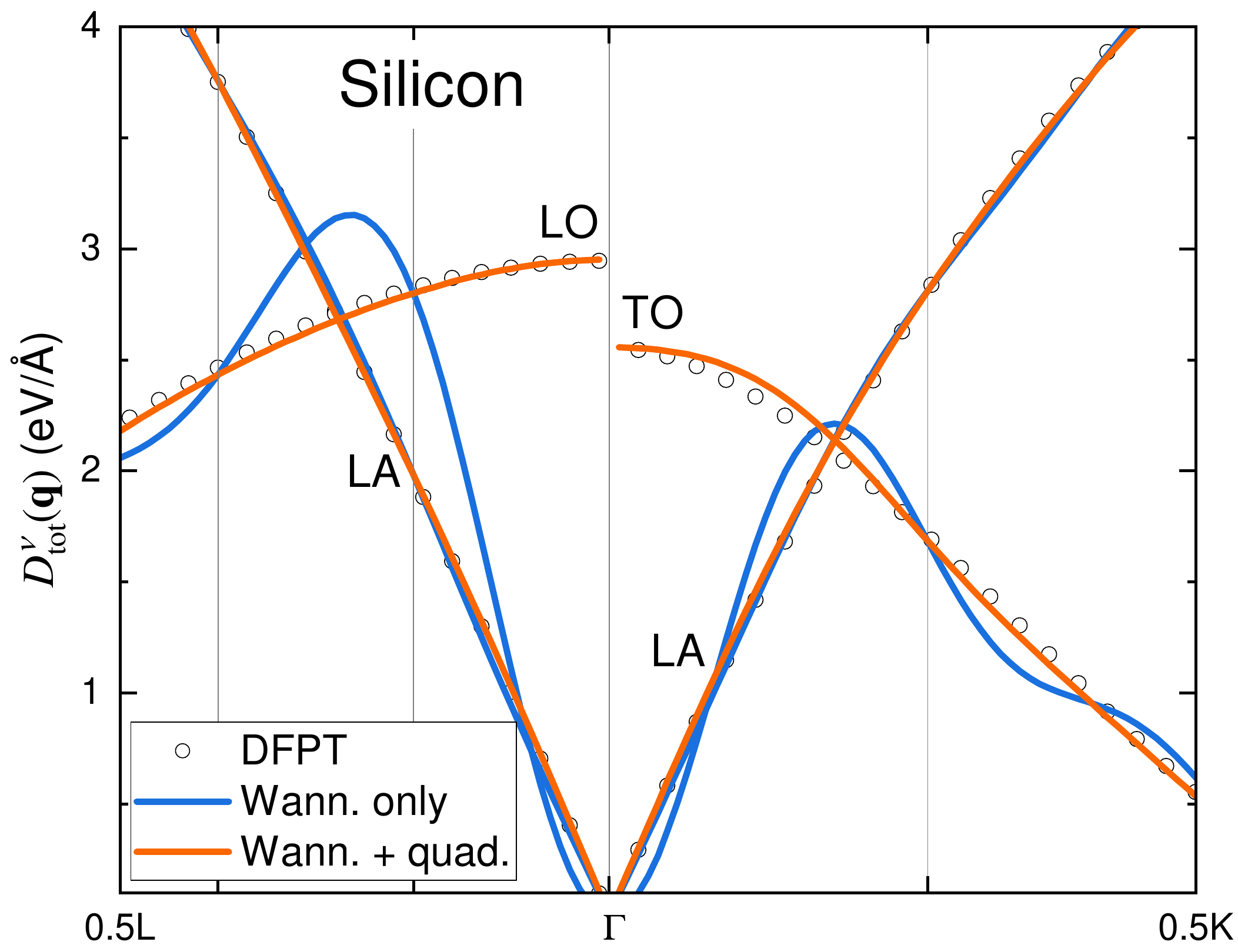}
\caption{Mode-resolved $e$-ph coupling strength [see Eq.~(\ref{eq:deformation_pot})] in silicon, computed using the lowest valence band. 
The electron momentum $\kk$ is fixed at the $\Gamma$ point and the phonon wave-vector $\qq$ is varied along high-symmetry lines in the Brillouin zone. 
Benchmark results from DFPT (black circles) are compared with Wannier interpolation with the quadrupole $e$-ph interaction included (orange line) or neglected (blue line). 
The coarse-grid $\qq$-points are indicated with vertical lines.
}\label{fig:silicon_g}
\end{figure}
% SILICON RESULTS
The long-range quadrupole $e$-ph interaction is present in a wide range of semiconductors and insulators, where the atomic dynamical quadrupoles are in general non-zero. 
%Metals are an exception given that long-range fields are effectively screened by the electron gas.   
% OPTICAL MODES SILICON
We illustrate this point by studying silicon, a simple nonpolar semiconductor in which the Born charges $-$ and thus the Fr\"ohlich interaction $-$ vanish and the presence of long-range interactions is not immediately obvious. 
Figure~\ref{fig:silicon_g} shows the $e$-ph coupling strength, $D_\text{tot}^\nu(\qq)$ in Eq.~(\ref{eq:deformation_pot}), computed directly using DFPT as a benchmark 
and compared with Wannier interpolation with and without inclusion of the quadrupole term. 
The DFPT benchmark $e$-ph matrix elements for optical modes approach a constant value as $\qq \rightarrow 0$, as we show for the LO mode in the $\Gamma-L$ direction 
and the transverse optical (TO) mode along $\Gamma-K$. This trend is distinctive of the quadrupole $e$-ph interaction, which is of order $q^0$ in the long-wavelength limit. 
\\
\indent 
If the quadrupole term is neglected and all $e$-ph interactions are treated as short-ranged, the $e$-ph matrix elements for optical modes in silicon incorrectly vanish as $\qq \rightarrow 0$. 
The interpolated values for optical modes are underestimated between the $\Gamma$ point, where the error is greatest, and its nearest-neighbor $\qq$-points in the coarse grid, where the error vanishes. 
Outside this $\qq$-point region close to $\Gamma$, the interpolated matrix elements without the quadrupole interaction still deviate from the DFPT result, 
although the error is smaller than near $\Gamma$. 
When the quadrupole term is included, the long-range $e$-ph interactions for the optical modes are captured correctly, as can be seen for the Wannier plus quadrupole curves in Fig.~\ref{fig:silicon_g}. 
The root-mean-square deviation of $D_\text{tot}^\nu(\qq)$ from DFPT, for the optical branches shown in Fig.~\ref{fig:silicon_g}, is 0.78~$\text{eV}/\text{\AA}$ when the quadrupole term is neglected versus 0.03~$\text{eV}/\text{\AA}$ when the quadrupole term is included in the interpolation. This result highlights the importance of the quadrupole term to correctly capture long-range $e$-ph interactions in nonpolar semiconductors.
\\
\indent
% ACOUSTIC MODES SILICON
Observe also how for acoustic modes in silicon the quadrupole term has a nearly negligible effect, as we show for the longitudinal acoustic (LA) mode in Fig.~\ref{fig:silicon_g}. 
As contracting the dynamical quadrupoles $\bf{Q}_\kappa$ with a rigid shift of the lattice leads to a vanishing quadrupole contribution~\cite{Vogl1976},  
one can obtain the quadrupole acoustic sum rule $\sum_{\alpha} Q_{\kappa,\alpha\beta\gamma}=0$ for nonpolar materials~\cite{Vogl1976}. 
This sum rule, which is satisfied by the dynamical quadrupole values we employ for silicon~\cite{Royo2019}, leads to a negligible quadrupole correction for acoustic modes in the long-wavelength limit.  
Though we focus on silicon in this work, on the basis of our results we expect sizable quadrupole contributions for optical modes, and negligible for acoustic modes, in all nonpolar semiconductors.
\\
\indent
\begin{figure}[!t]
\centering 
\includegraphics[width=0.95\columnwidth]{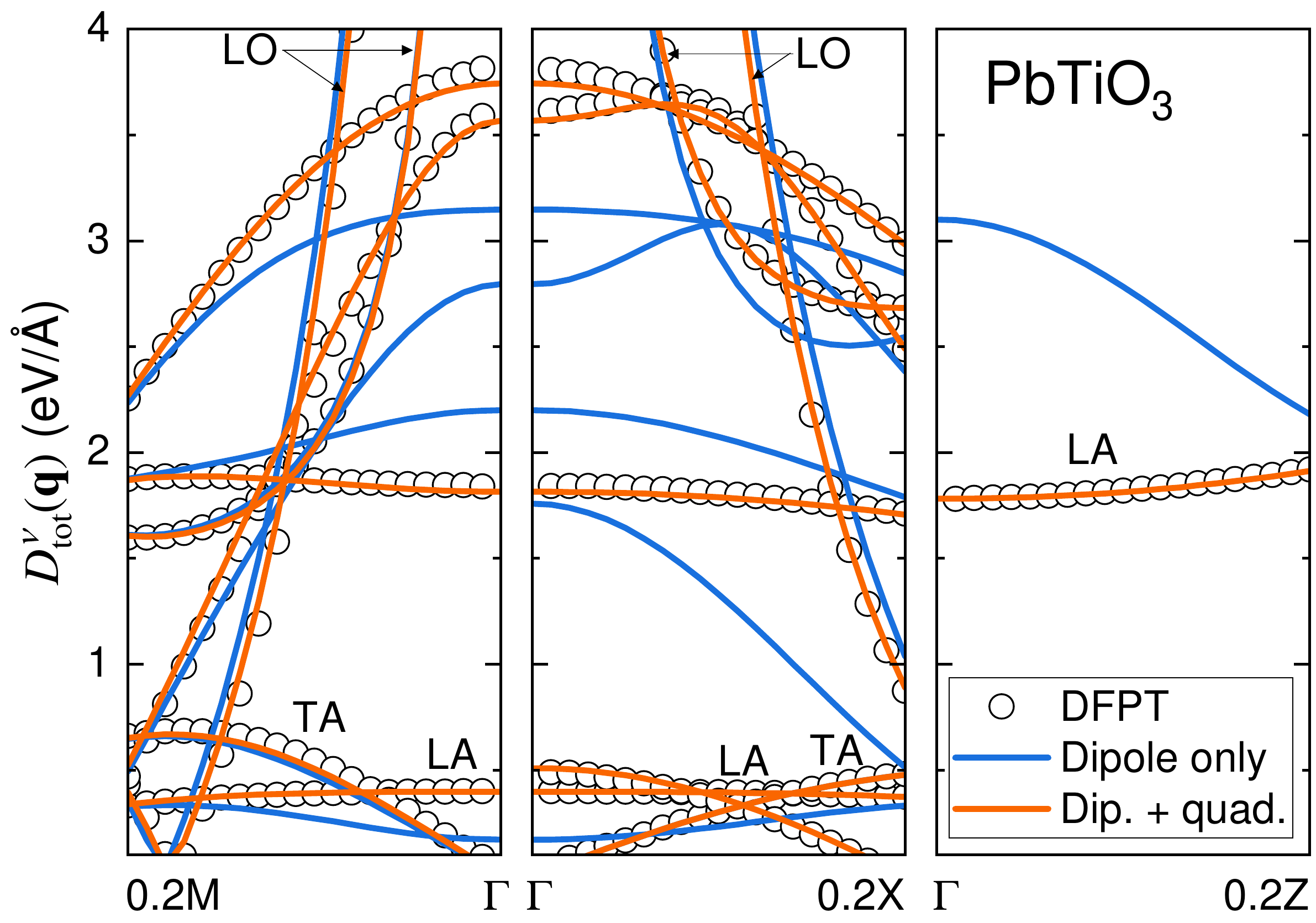}
\caption{Mode-resolved $e$-ph coupling strength [see Eq.~(\ref{eq:deformation_pot})] in tetragonal $\text{PbTiO}_3$, computed using the lowest conduction band. 
The initial electron momentum is fixed at the $\Gamma$ point and the phonon wave-vector $\qq$ is varied along high-symmetry lines in the Brillouin zone.
Benchmark results from DFPT (black circles) are compared with Wannier interpolation plus the Fr\"ohlich interaction (blue line) and Wannier interpolation plus the Fr\"ohlich and quadrupole interactions (orange line). 
}\label{fig:pbtio_g}
\end{figure}
%
% PbTiO3 E-PH RESULTS
%
The quadrupole $e$-ph interaction is particularly critical in piezoelectric materials, as discussed here for tetragonal $\text{PbTiO}_3$, a prototypical piezoelectric insulator. 
Piezoelectric materials are polar noncentrosymmetric systems with non-zero Born charges.  
As a result, the dipole Fr\"ohlich interaction is dominant for LO modes near $\qq \rightarrow 0$ due to its $1/q$ divergence. The quadrupole contribution is expected to be important for TO and acoustic modes 
(the quadrupole acoustic sum rule does not hold for polar noncentrosymmetric crystals).
\\
\indent
Figure~\ref{fig:pbtio_g} shows the $e$-ph coupling strength, $D_\text{tot}^\nu(\qq)$ in Eq.~(\ref{eq:deformation_pot}), for the DFPT benchmark in tetragonal $\text{PbTiO}_3$, 
and compares it with interpolated results that include only the Fr\"ohlich dipole interaction or both the Fr\"ohlich and the quadrupole interactions. 
The short-range interactions are included through Wannier interpolation in both cases. 
%We focus on the region near $\qq = 0$ where the long-range interactions are dominant. 
%
When only the Fr\"ohlich dipole interaction is included, the $e$-ph matrix elements deviate dramatically from the DFPT results. 
The values are either overestimated or underestimated depending on the phonon mode considered, 
with deviations from DFPT that depend strongly on the direction in which $\qq$ approaches $\Gamma$ due to the non-analytic character of the long-range $e$-ph interactions. 
When the quadrupole $e$-ph interaction is taken into account, the interpolated $e$-ph coupling strength matches the DFPT result very accurately for \emph{all} phonon modes. 
For LO modes, the quadrupole correction is moderate due to the dominant Fr\"ohlich term near $\qq \!=\! 0$. For other optical and acoustic modes with a finite $e$-ph coupling at $\qq \!=\! 0$, 
the quadrupole term removes the large error in the dipole-only results (up to an order of magnitude) and gives $e$-ph matrix elements in nearly exact agreement with DFPT.  
For the branches shown in Fig.~\ref{fig:pbtio_g}, the root-mean-square deviation of $D_\text{tot}^\nu(\qq)$ from DFPT is 0.46~$\text{eV}/\text{\AA}$ for dipole-only results versus 0.03~$\text{eV}/\text{\AA}$ for our dipole plus quadrupole interpolation scheme. It is clear that the quadrupole term is essential in piezoelectric materials for all phonon modes. %except the LO.
\\
\indent 
% PIEZOELECTRIC E-PH INTERACTION
Contrary to silicon and nonpolar materials, the quadrupole term has a large effect for acoustic modes in piezoelectric materials, 
where it is one of the two contributions to the so-called piezoelectric $e$-ph interaction~\cite{mahanCondensed2011}. 
%  
%Both the dipole and quadrupole terms contribute to the piezoelectric $e$-ph interaction due to the acoustic modes~\cite{Vogl1976}. 
Expanding the phonon eigenvectors at $\qq \!\rightarrow\! 0$ as $\mathbf{e}_{\nu\qq} \!\approx\! \mathbf{e}^{(0)}_{\nu\qq} + i\mathbf{q}\cdot\mathbf{e}^{(1)}_{\nu\qq}$,  
one finds two contributions of order $q^0$~\cite{Vogl1976}. %to the PE $e$-ph interaction 
One is from the Born charges, $\mathbf{Z}_{\kappa} \mathbf{e}^{(1)}_{\nu \qq}$, and is a dipole-like interaction generated by atoms with a net charge experiencing different displacements due to strain from an acoustic mode. 
The other is from the dynamical quadrupoles, $\mathbf{Q}_{\kappa} \mathbf{e}^{(0)}_{\nu\qq}$, and is associated with a clamped-ion electronic polarization~\cite{Bernardini1997}. 
The \textit{ab initio} Fr\"ohlich interaction includes only the former term, namely the strain component of the piezoelectric $e$-ph interaction, and thus the dipole-only scheme leads to large errors for acoustic phonons in PbTiO$_3$ (see Fig.~\ref{fig:pbtio_g}) 
as it neglects the important electronic quadrupole contribution. 
Until now, the \textit{ab initio} Fr\"ohlich term has been mistakenly thought to fully capture piezoelectric $e$-ph interactions. Our results demonstrate that both dipole and quadrupole terms are essential for accurate acoustic mode $e$-ph interactions in piezoelectric materials~\cite{saghi-szaboFirstPrinciples1998}. 
The relative magnitude of the strain and quadrupole contributions is material dependent $-$ the two terms can nearly cancel each other out, as we have shown elsewhere for GaN~\cite{Jhalani2020}, 
or their ratio can be mode and phonon wave-vector dependent, as we find in PbTiO$_3$.  
\\
\indent %relative magnitude

\subsection{Quadrupole contribution to the scattering rate} 
%
% ------- FIGURE 4:  SCATTERING RATES -----------
%
\begin{figure}[!t]
\centering 
\includegraphics[width=0.95\columnwidth]{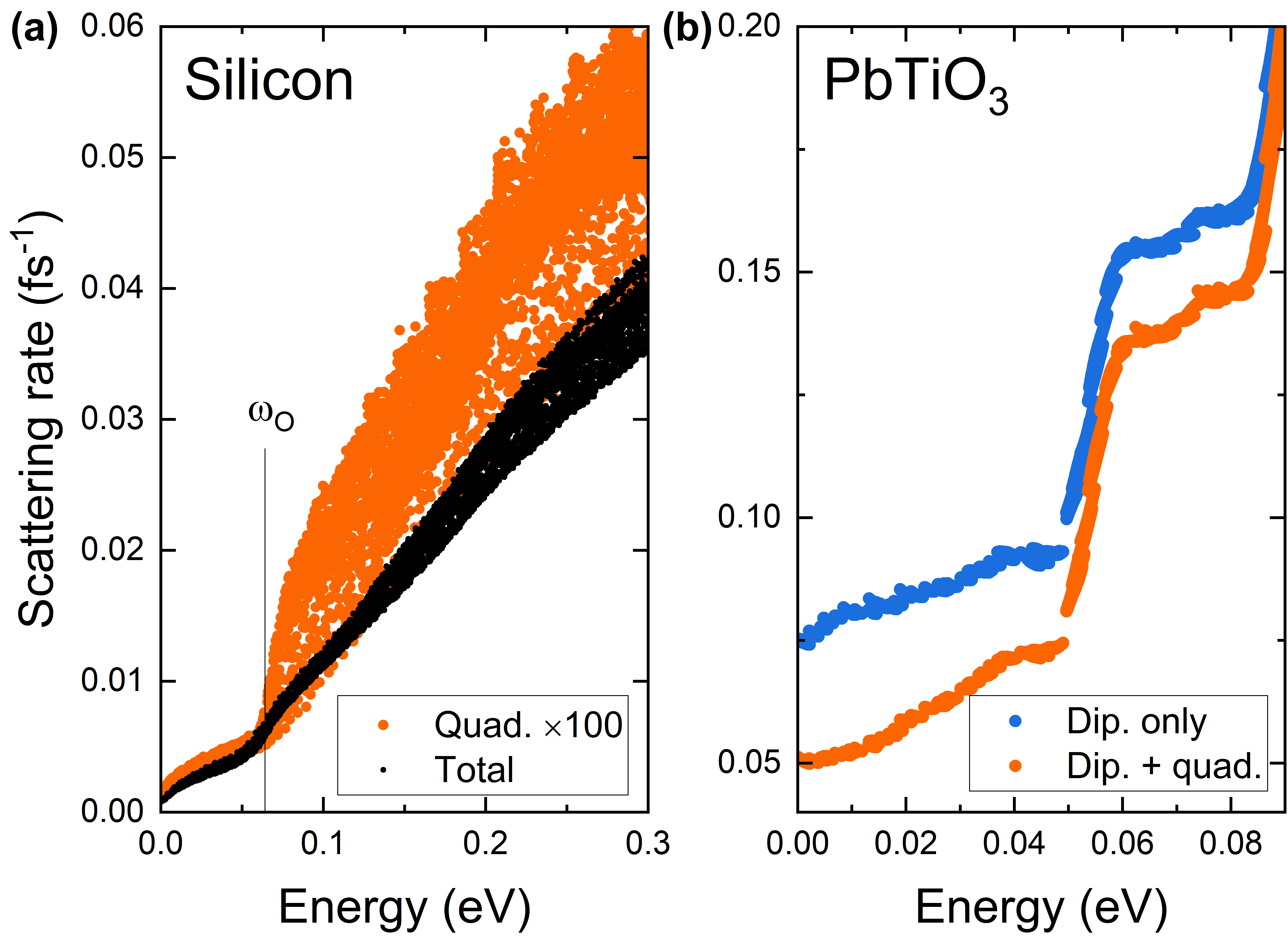}
\caption{Room temperature scattering rate versus electron energy (referenced to the conduction band minimum) in (a) silicon and (b) $\text{PbTiO}_3$. 
For silicon, we plot the quadrupole contribution multiplied by 100 (orange) and the total scattering rate (black), which includes the short-range and the quadrupole contributions. 
For $\text{PbTiO}_3$, we show the long-range scattering rate computed using only the Fr\"ohlich interaction (blue) or both the Fr\"ohlich and quadrupole interactions (orange). 
}\label{fig:rate}
\end{figure} 
Because the quadrupole interaction has a significant effect on the $e$-ph matrix elements, we expect that it also plays a role in calculations of the $e$-ph scattering rate and mobility. 
Figure~\ref{fig:rate}(a) shows both the quadrupole contribution and the total $e$-ph scattering rate in silicon at 300~K for electron energies near the conduction band minimum. 
We find that the quadrupole contribution to the scattering rate is about 1\% of the total scattering rate at temperatures between 100$-$400~K. 
At electron energies below the optical phonon emission threshold in silicon ($\hbar \omega_{\rm O}\approx65$~meV relative to the conduction band minimum), 
absorption and emission of acoustic phonons dominate the scattering processes, and thus we find a small correction due to the quadrupole interaction, which minimally affects acoustic modes in silicon. 
Since the quadrupole acoustic sum rule holds only in the long wavelength limit, the quadrupole interaction can still contribute to finite-$\qq$ acoustic scattering, 
as is shown by the fact that the quadrupole scattering rate at energy below $\hbar \omega_{\rm O}$ is proportional to the total scattering rate. 
The quadrupole contribution increases sharply above the optical emission threshold because the quadrupole term is greater for optical modes in silicon.  
For the same reason, the relative contribution of the quadrupole term increases slightly with temperature in the 100$-$400~K range, varying from 1\% of the total scattering rate at 100~K to 1.5\% at 400~K. 
\\
\indent
%
% PbTiO3
% 
The effect of the quadrupole interaction on the scattering rates is greater in $\text{PbTiO}_3$. Our analysis focuses on the $e$-ph scattering rate due to the long-range $e$-ph interactions, although similar conclusions hold for the total scattering rate. 
Figure~\ref{fig:rate}(b) shows the long-range $e$-ph scattering rate in $\text{PbTiO}_3$ at 300~K as a function of electron energy, 
comparing results that include only the dipole Fr\"ohlich interaction with results from our approach including both the dipole and quadrupole terms.
The scattering rate from the long-range $e$-ph interactions is lower at all energies when the quadrupole term is taken into account. 
The difference is greatest near the band edge, where the scattering rate due to the dipole interaction alone is 0.075 fs$^{-1}$ versus a 50\% smaller value of 0.050 fs$^{-1}$ for dipole plus quadrupole.
\\
\indent
These trends can be understood on the basis of the $e$-ph matrix element analysis in Fig.~\ref{fig:pbtio_g}. 
The errors found when neglecting the quadrupole term in $D_\text{tot}^\nu(\qq)$, which is proportional to the absolute value of the matrix elements [see Eq.~(\ref{eq:deformation_pot})], 
are amplified in calculations of the scattering rate, which is proportional to the square of the matrix elements. 
The largest errors we find for $D_\text{tot}^\nu(\qq)$ are in the $\qq \rightarrow 0$ limit, especially for the acoustic modes. 
For example, for the LA mode in the $\Gamma-M$ and $\Gamma-X$ directions, the value of $D_\text{tot}^\nu(\qq)$ from the dipole-only calculation is 0.17~$\text{eV}/\text{\AA}$ 
compared to a twice-greater value of 0.40~$\text{eV}/\text{\AA}$ when the quadrupole term is included. This leads to a four-fold increase of the LA mode scattering rate upon including the quadrupole interaction. 
Opposite to the silicon case, in PbTiO$_3$ the relative magnitude of the quadrupole correction is greater at lower temperatures because the quadrupole interaction is stronger for acoustic modes. 
Near the band edge, we find quadrupole corrections to the long-range scattering rate ranging from 97\% at 100~K to 38\% at 400~K. 
Given that low-energy electronic states near the band edge govern transport properties, including the quadrupole term  
is critical to accurately computing electronic transport.
\\
\indent
\subsection{Quadrupole contribution to the mobility}
%
% ------- FIGURE 5: MOBILITY ----------
%
\begin{figure}[!t]
\centering 
\includegraphics[width=1.0\columnwidth]{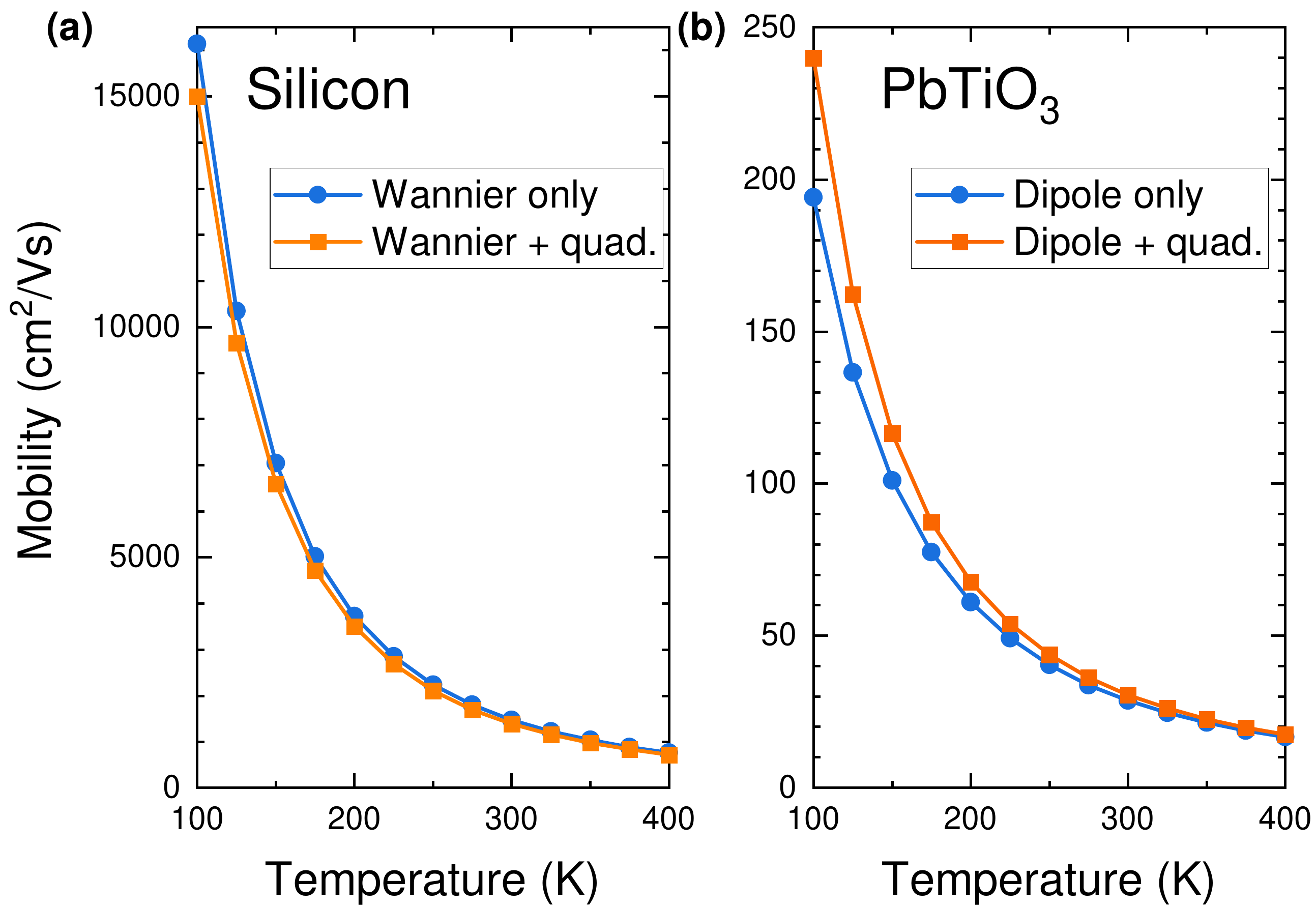}
\caption{Computed temperature-dependent electron mobility in (a) silicon and (b) tetragonal $\text{PbTiO}_3$.  
The plot compares the mobility obtained when the quadrupole $e$-ph interaction is included (orange squares) or neglected (blue circles). 
The $\text{PbTiO}_3$ results are for transport in the basal $xy$ plane. 
}\label{fig:mobility}
\end{figure}
The effect of the quadrupole $e$-ph interaction on the mobility is noteworthy. 
Figure~\ref{fig:mobility}(a) shows the temperature dependent electron mobility in silicon computed with and without the quadrupole term.  
Including the quadrupole interaction reduces the computed mobility by approximately 5$-$10\% due to the increased $e$-ph coupling strength and scattering rates. 
For example, the computed mobility at 300~K is 1390~$\text{cm}^2/\text{Vs}$ when including the quadrupole interaction 
versus a value of 1473~$\text{cm}^2/\text{Vs}$ with the conventional interpolation approach in which all $e$-ph interactions in silicon are treated as short-ranged.
This discrepancy is due to the underestimation of the $e$-ph coupling strength for optical modes in the conventional approach, especially at small values of $\qq$ as shown in Fig.~\ref{fig:silicon_g}. 
As a result, intravalley scattering due to optical modes is underestimated without the quadrupole term, leading to an artificially high mobility. 
Although we focus on silicon, we expect that these trends apply in general to nonpolar semiconductors because small-$\qq$ optical $e$-ph coupling will consistently be underestimated without the quadrupole term. 
The long-range quadrupole $e$-ph interaction is thus surprisingly manifest in the transport properties of nonpolar materials. 
\\
\indent
We find an opposite trend in $\text{PbTiO}_3$, in which including the quadrupole interaction increases the mobility by 10$-$25\% between 100$-$400 K, as seen in Fig.~\ref{fig:mobility}(b). 
The quadrupole term gives a larger correction at lower temperatures, reaching values up to $\sim$25\% at 100~K. 
This result is due to the dominant acoustic mode contribution at low temperatures together with the large quadrupole correction for acoustic modes in piezoelectric materials. 
At higher temperatures, where optical mode scattering is dominant and acoustic scattering less important, the quadrupole contribution is smaller, only about 10\% at 400~K. 
Due to differences in the quadrupole interaction for different phonon modes and to varying mode contributions to the mobility as a function of temperature, 
including the quadrupole term corrects the temperature dependence of the mobility~\cite{Jhalani2020} and is essential in piezoelectric materials.
\\
\indent
%
%  SECTION: -------------   DISCUSSION   -----------------
%
\section{Discussion}  
\vspace{-10pt}
We briefly discuss a technical aspect of the $e$-ph matrix element interpolation. 
The treatment of long-wavelength perturbations with wave-vector $\qq \rightarrow 0$ in DFPT is critical in semiconductors and insulators~\cite{calandraAdiabatic2010,Agapito2018}. 
The lattice-periodic part of the phonon perturbation potential, $\Delta v_{\qq}(\rr)$, is the sum of a Coulomb and an exchange-correlation contribution,
\begin{equation}
\Delta v_{\qq}(\rr)= \Delta v_{\qq,C}(\rr) +  \Delta v_{\qq,X\!C}(\rr).
\end{equation}
The Coulomb contribution $\Delta v_{\qq,C}(\rr)$ combines the variation of the Hartree and electron-nuclei interactions. 
Its integral over the unit cell~\cite{calandraAdiabatic2010},
\begin{equation}
\Delta(\qq)=\frac{1}{\Omega}\int_\Omega {d\rr ~ v_{\qq,C}(\rr) },
\end{equation}
is well-behaved for insulators (and semiconductors) at finite $\qq$ values, but is ill-defined at $\qq = 0$. %and discontinuous 
First-principles codes such as {\sc Quantum ESPRESSO}~\cite{giannozziQUANTUM2009} subtract $\Delta(\qq)$ from the perturbation potential at $\qq\!=\!0$, thus making it discontinuous at $\qq=0$. Therefore, due to both the discontinuity at $\qq=0$ and the non-analytic behavior near $\qq=0$, the $e$-ph matrix elements are challenging to interpolate in the long-wavelength limit. 
\\ 
\indent
In our scheme, we identify the quadrupole interaction as the key long-range term in nonpolar materials, and remove the non-analytic behavior near $\qq=0$ on the coarse grid by subtracting the quadrupole term. 
This strategy improves the interpolation near $\qq=0$ in nonpolar materials, at once capturing the correct physics and smoothing the coarse-grid matrix element to be interpolated. 
Due to the non-analytic behavior, denser DFPT grids cannot fully remove the interpolation error if the quadrupole term is not subtracted on the coarse grid~\cite{Brunin2020}.  
For polar materials such as $\text{PbTiO}_3$, the non-analytic behavior is due to both the dipole (Fr\"ohlich) and quadrupole long-range $e$-ph interactions. 
By subtracting both terms in our scheme in polar materials, the coarse-grid matrix elements to be interpolated are made smooth and the interpolation approach more reliable. 
The non-analytic behavior of the Coulomb potential is correctly reconstructed by adding back the dipole (in polar materials) and quadrupole (in all insulators) contributions after interpolation. 
% 
% SECTION: ----------------   CONCLUSION   ----------------------
%
\section{Conclusion} 
\vspace{-10pt}
In summary, we developed an accurate approach for computing the quadrupole $e$-ph interaction from first principles. 
This advance resolves the outstanding problem of correctly quantifying long-range $e$-ph interactions for all phonon modes in semiconductors and insulators.  
Our results clearly show that the quadrupole interactions are crucial for obtaining accurate $e$-ph matrix elements, scattering rates and electronic transport properties. 
The quadrupole effect is particularly apparent in piezoelectric materials such as wurtzite GaN~\cite{Jhalani2020} and PbTiO$_3$, in which neglecting the quadrupole interaction leads to large and uncontrolled errors. 
The method introduced in this work enables accurate calculations of electrical transport, thermoelectric properties and superconductivity in a wide range of materials.
\begin{acknowledgments}
\vspace{-10pt}
\noindent
J.P. acknowledges support by the Korea Foundation for Advanced Studies. 
V.J. thanks the Resnick Sustainability Institute at Caltech for fellowship support. 
This work was supported by the National Science Foundation under Grants No. DMR-1750613 for theory development and ACI-1642443 for code development. 
J.-J.Z. acknowledges partial support from the Joint Center for Artificial Photosynthesis, a DOE Energy Innovation Hub, as follows: the development of some computational methods employed in this work was supported through the Office of Science of the U.S. Department of Energy under Award No. DE-SC0004993. 
C.E.D. acknowledges support from the National Science Foundation under Grant No. DMR-1918455. 
The Flatiron Institute is a division of the Simons Foundation.
This research used resources of the National Energy Research Scientific Computing Center, a DOE Office of Science User Facility supported by the Office of Science of the U.S. Department of Energy under Contract No. DE-AC02-05CH11231.\\
\end{acknowledgments}
\textit{Note added.}$-$ While writing this manuscript, we became aware of a related preprint by another group~\cite{Brunin2020}. 
Their article analyzes how the quadrupole term improves $e$-ph matrix element interpolation, while this work focuses on the physics of $e$-ph interactions and the quadrupole interaction for different phonon modes and materials. 
%
%  REFERENCES
%
\bibliographystyle{apsrev4-1}

\begin{thebibliography}{37}%
\makeatletter
\providecommand \@ifxundefined [1]{%
 \@ifx{#1\undefined}
}%
\providecommand \@ifnum [1]{%
 \ifnum #1\expandafter \@firstoftwo
 \else \expandafter \@secondoftwo
 \fi
}%
\providecommand \@ifx [1]{%
 \ifx #1\expandafter \@firstoftwo
 \else \expandafter \@secondoftwo
 \fi
}%
\providecommand \natexlab [1]{#1}%
\providecommand \enquote  [1]{``#1''}%
\providecommand \bibnamefont  [1]{#1}%
\providecommand \bibfnamefont [1]{#1}%
\providecommand \citenamefont [1]{#1}%
\providecommand \href@noop [0]{\@secondoftwo}%
\providecommand \href [0]{\begingroup \@sanitize@url \@href}%
\providecommand \@href[1]{\@@startlink{#1}\@@href}%
\providecommand \@@href[1]{\endgroup#1\@@endlink}%
\providecommand \@sanitize@url [0]{\catcode `\\12\catcode `\$12\catcode
  `\&12\catcode `\#12\catcode `\^12\catcode `\_12\catcode `\%12\relax}%
\providecommand \@@startlink[1]{}%
\providecommand \@@endlink[0]{}%
\providecommand \url  [0]{\begingroup\@sanitize@url \@url }%
\providecommand \@url [1]{\endgroup\@href {#1}{\urlprefix }}%
\providecommand \urlprefix  [0]{URL }%
\providecommand \Eprint [0]{\href }%
\providecommand \doibase [0]{http://dx.doi.org/}%
\providecommand \selectlanguage [0]{\@gobble}%
\providecommand \bibinfo  [0]{\@secondoftwo}%
\providecommand \bibfield  [0]{\@secondoftwo}%
\providecommand \translation [1]{[#1]}%
\providecommand \BibitemOpen [0]{}%
\providecommand \bibitemStop [0]{}%
\providecommand \bibitemNoStop [0]{.\EOS\space}%
\providecommand \EOS [0]{\spacefactor3000\relax}%
\providecommand \BibitemShut  [1]{\csname bibitem#1\endcsname}%
\let\auto@bib@innerbib\@empty
%</preamble>
\bibitem [{\citenamefont {Ziman}(2001)}]{zimanElectrons2001}%
  \BibitemOpen
  \bibfield  {author} {\bibinfo {author} {\bibfnamefont {J.~M.}\ \bibnamefont
  {Ziman}},\ }\href@noop {} {\emph {\bibinfo {title} {Electrons and Phonons:
  The Theory of Transport Phenomena in Solids}}}\ (\bibinfo  {publisher}
  {{Oxford University Press}},\ \bibinfo {year} {2001})\BibitemShut {NoStop}%
\bibitem [{\citenamefont {Bernardi}(2016)}]{Bernardi2016}%
  \BibitemOpen
  \bibfield  {author} {\bibinfo {author} {\bibfnamefont {M.}~\bibnamefont
  {Bernardi}},\ }\href {\doibase 10.1140/epjb/e2016-70399-4} {\bibfield
  {journal} {\bibinfo  {journal} {Eur. Phys. J. B}\ }\textbf {\bibinfo {volume}
  {89}},\ \bibinfo {pages} {239} (\bibinfo {year} {2016})}\BibitemShut
  {NoStop}%
\bibitem [{\citenamefont {Bernardi}\ \emph {et~al.}(2014)\citenamefont
  {Bernardi}, \citenamefont {{Vigil-Fowler}}, \citenamefont {Lischner},
  \citenamefont {Neaton},\ and\ \citenamefont {Louie}}]{bernardiInitio2014}%
  \BibitemOpen
  \bibfield  {author} {\bibinfo {author} {\bibfnamefont {M.}~\bibnamefont
  {Bernardi}}, \bibinfo {author} {\bibfnamefont {D.}~\bibnamefont
  {{Vigil-Fowler}}}, \bibinfo {author} {\bibfnamefont {J.}~\bibnamefont
  {Lischner}}, \bibinfo {author} {\bibfnamefont {J.~B.}\ \bibnamefont
  {Neaton}}, \ and\ \bibinfo {author} {\bibfnamefont {S.~G.}\ \bibnamefont
  {Louie}},\ }\href {\doibase 10.1103/PhysRevLett.112.257402} {\bibfield
  {journal} {\bibinfo  {journal} {Phys. Rev. Lett.}\ }\textbf {\bibinfo
  {volume} {112}},\ \bibinfo {pages} {257402} (\bibinfo {year}
  {2014})}\BibitemShut {NoStop}%
\bibitem [{\citenamefont {Zhou}\ and\ \citenamefont
  {Bernardi}(2016)}]{Zhou2016}%
  \BibitemOpen
  \bibfield  {author} {\bibinfo {author} {\bibfnamefont {J.-J.}\ \bibnamefont
  {Zhou}}\ and\ \bibinfo {author} {\bibfnamefont {M.}~\bibnamefont
  {Bernardi}},\ }\href {\doibase 10.1103/PhysRevB.94.201201} {\bibfield
  {journal} {\bibinfo  {journal} {Phys. Rev. B}\ }\textbf {\bibinfo {volume}
  {94}},\ \bibinfo {pages} {201201(R)} (\bibinfo {year} {2016})}\BibitemShut
  {NoStop}%
\bibitem [{\citenamefont {Jhalani}\ \emph {et~al.}(2017)\citenamefont
  {Jhalani}, \citenamefont {Zhou},\ and\ \citenamefont
  {Bernardi}}]{jhalaniUltrafast2017}%
  \BibitemOpen
  \bibfield  {author} {\bibinfo {author} {\bibfnamefont {V.~A.}\ \bibnamefont
  {Jhalani}}, \bibinfo {author} {\bibfnamefont {J.-J.}\ \bibnamefont {Zhou}}, \
  and\ \bibinfo {author} {\bibfnamefont {M.}~\bibnamefont {Bernardi}},\ }\href
  {\doibase 10.1021/acs.nanolett.7b02212} {\bibfield  {journal} {\bibinfo
  {journal} {Nano Lett.}\ }\textbf {\bibinfo {volume} {17}},\ \bibinfo {pages}
  {5012} (\bibinfo {year} {2017})}\BibitemShut {NoStop}%
\bibitem [{\citenamefont {Lee}\ \emph {et~al.}(2018)\citenamefont {Lee},
  \citenamefont {Zhou}, \citenamefont {Agapito},\ and\ \citenamefont
  {Bernardi}}]{leeCharge2018}%
  \BibitemOpen
  \bibfield  {author} {\bibinfo {author} {\bibfnamefont {N.-E.}\ \bibnamefont
  {Lee}}, \bibinfo {author} {\bibfnamefont {J.-J.}\ \bibnamefont {Zhou}},
  \bibinfo {author} {\bibfnamefont {L.~A.}\ \bibnamefont {Agapito}}, \ and\
  \bibinfo {author} {\bibfnamefont {M.}~\bibnamefont {Bernardi}},\ }\href
  {\doibase 10.1103/PhysRevB.97.115203} {\bibfield  {journal} {\bibinfo
  {journal} {Phys. Rev. B}\ }\textbf {\bibinfo {volume} {97}},\ \bibinfo
  {pages} {115203} (\bibinfo {year} {2018})}\BibitemShut {NoStop}%
\bibitem [{\citenamefont {Zhou}\ \emph {et~al.}(2018)\citenamefont {Zhou},
  \citenamefont {Hellman},\ and\ \citenamefont {Bernardi}}]{Zhou2018}%
  \BibitemOpen
  \bibfield  {author} {\bibinfo {author} {\bibfnamefont {J.-J.}\ \bibnamefont
  {Zhou}}, \bibinfo {author} {\bibfnamefont {O.}~\bibnamefont {Hellman}}, \
  and\ \bibinfo {author} {\bibfnamefont {M.}~\bibnamefont {Bernardi}},\ }\href
  {\doibase 10.1103/PhysRevLett.121.226603} {\bibfield  {journal} {\bibinfo
  {journal} {Phys. Rev. Lett.}\ }\textbf {\bibinfo {volume} {121}},\ \bibinfo
  {pages} {226603} (\bibinfo {year} {2018})}\BibitemShut {NoStop}%
\bibitem [{\citenamefont {Zhou}\ and\ \citenamefont
  {Bernardi}(2019)}]{zhouPredicting2019}%
  \BibitemOpen
  \bibfield  {author} {\bibinfo {author} {\bibfnamefont {J.-J.}\ \bibnamefont
  {Zhou}}\ and\ \bibinfo {author} {\bibfnamefont {M.}~\bibnamefont
  {Bernardi}},\ }\href {\doibase 10.1103/PhysRevResearch.1.033138} {\bibfield
  {journal} {\bibinfo  {journal} {Phys. Rev. Research}\ }\textbf {\bibinfo
  {volume} {1}},\ \bibinfo {pages} {033138} (\bibinfo {year}
  {2019})}\BibitemShut {NoStop}%
\bibitem [{\citenamefont {Zhou}\ \emph {et~al.}()\citenamefont {Zhou},
  \citenamefont {Park}, \citenamefont {Lu}, \citenamefont {Maliyov},
  \citenamefont {Tong},\ and\ \citenamefont {Bernardi}}]{Perturbo}%
  \BibitemOpen
  \bibfield  {author} {\bibinfo {author} {\bibfnamefont {J.-J.}\ \bibnamefont
  {Zhou}}, \bibinfo {author} {\bibfnamefont {J.}~\bibnamefont {Park}}, \bibinfo
  {author} {\bibfnamefont {I.-T.}\ \bibnamefont {Lu}}, \bibinfo {author}
  {\bibfnamefont {I.}~\bibnamefont {Maliyov}}, \bibinfo {author} {\bibfnamefont
  {X.}~\bibnamefont {Tong}}, \ and\ \bibinfo {author} {\bibfnamefont
  {M.}~\bibnamefont {Bernardi}},\ }\href@noop {} {\ }\Eprint
  {http://arxiv.org/abs/2002.02045} {arXiv:2002.02045} \BibitemShut {NoStop}%
\bibitem [{\citenamefont {Li}(2015)}]{liElectrical2015}%
  \BibitemOpen
  \bibfield  {author} {\bibinfo {author} {\bibfnamefont {W.}~\bibnamefont
  {Li}},\ }\href {\doibase 10.1103/PhysRevB.92.075405} {\bibfield  {journal}
  {\bibinfo  {journal} {Phys. Rev. B}\ }\textbf {\bibinfo {volume} {92}},\
  \bibinfo {pages} {075405} (\bibinfo {year} {2015})}\BibitemShut {NoStop}%
\bibitem [{\citenamefont {Liu}\ \emph {et~al.}(2017)\citenamefont {Liu},
  \citenamefont {Zhou}, \citenamefont {Liao}, \citenamefont {Singh},\ and\
  \citenamefont {Chen}}]{liuFirstprinciples2017}%
  \BibitemOpen
  \bibfield  {author} {\bibinfo {author} {\bibfnamefont {T.-H.}\ \bibnamefont
  {Liu}}, \bibinfo {author} {\bibfnamefont {J.}~\bibnamefont {Zhou}}, \bibinfo
  {author} {\bibfnamefont {B.}~\bibnamefont {Liao}}, \bibinfo {author}
  {\bibfnamefont {D.~J.}\ \bibnamefont {Singh}}, \ and\ \bibinfo {author}
  {\bibfnamefont {G.}~\bibnamefont {Chen}},\ }\href {\doibase
  10.1103/PhysRevB.95.075206} {\bibfield  {journal} {\bibinfo  {journal} {Phys.
  Rev. B}\ }\textbf {\bibinfo {volume} {95}},\ \bibinfo {pages} {075206}
  (\bibinfo {year} {2017})}\BibitemShut {NoStop}%
\bibitem [{\citenamefont {Ma}\ \emph {et~al.}(2018)\citenamefont {Ma},
  \citenamefont {Nissimagoudar},\ and\ \citenamefont
  {Li}}]{maFirstprinciples2018}%
  \BibitemOpen
  \bibfield  {author} {\bibinfo {author} {\bibfnamefont {J.}~\bibnamefont
  {Ma}}, \bibinfo {author} {\bibfnamefont {A.~S.}\ \bibnamefont
  {Nissimagoudar}}, \ and\ \bibinfo {author} {\bibfnamefont {W.}~\bibnamefont
  {Li}},\ }\href {\doibase 10.1103/PhysRevB.97.045201} {\bibfield  {journal}
  {\bibinfo  {journal} {Phys. Rev. B}\ }\textbf {\bibinfo {volume} {97}},\
  \bibinfo {pages} {045201} (\bibinfo {year} {2018})}\BibitemShut {NoStop}%
\bibitem [{\citenamefont {Sohier}\ \emph {et~al.}(2018)\citenamefont {Sohier},
  \citenamefont {Campi}, \citenamefont {Marzari},\ and\ \citenamefont
  {Gibertini}}]{sohierMobility2018}%
  \BibitemOpen
  \bibfield  {author} {\bibinfo {author} {\bibfnamefont {T.}~\bibnamefont
  {Sohier}}, \bibinfo {author} {\bibfnamefont {D.}~\bibnamefont {Campi}},
  \bibinfo {author} {\bibfnamefont {N.}~\bibnamefont {Marzari}}, \ and\
  \bibinfo {author} {\bibfnamefont {M.}~\bibnamefont {Gibertini}},\ }\href
  {\doibase 10.1103/PhysRevMaterials.2.114010} {\bibfield  {journal} {\bibinfo
  {journal} {Phys. Rev. Materials}\ }\textbf {\bibinfo {volume} {2}},\ \bibinfo
  {pages} {114010} (\bibinfo {year} {2018})}\BibitemShut {NoStop}%
\bibitem [{\citenamefont {Martin}(2004)}]{martinElectronic2004}%
  \BibitemOpen
  \bibfield  {author} {\bibinfo {author} {\bibfnamefont {R.~M.}\ \bibnamefont
  {Martin}},\ }\href@noop {} {\emph {\bibinfo {title} {Electronic Structure:
  Basic Theory and Practical Methods}}}\ (\bibinfo  {publisher} {{Cambridge
  University Press}},\ \bibinfo {address} {{Cambridge, UK ; New York}},\
  \bibinfo {year} {2004})\BibitemShut {NoStop}%
\bibitem [{\citenamefont {Baroni}\ \emph {et~al.}(2001)\citenamefont {Baroni},
  \citenamefont {{de Gironcoli}}, \citenamefont {Dal~Corso},\ and\
  \citenamefont {Giannozzi}}]{Baroni-DFPT}%
  \BibitemOpen
  \bibfield  {author} {\bibinfo {author} {\bibfnamefont {S.}~\bibnamefont
  {Baroni}}, \bibinfo {author} {\bibfnamefont {S.}~\bibnamefont {{de
  Gironcoli}}}, \bibinfo {author} {\bibfnamefont {A.}~\bibnamefont
  {Dal~Corso}}, \ and\ \bibinfo {author} {\bibfnamefont {P.}~\bibnamefont
  {Giannozzi}},\ }\href {\doibase 10.1103/RevModPhys.73.515} {\bibfield
  {journal} {\bibinfo  {journal} {Rev. Mod. Phys.}\ }\textbf {\bibinfo {volume}
  {73}},\ \bibinfo {pages} {515} (\bibinfo {year} {2001})}\BibitemShut
  {NoStop}%
\bibitem [{\citenamefont {Agapito}\ and\ \citenamefont
  {Bernardi}(2018)}]{Agapito2018}%
  \BibitemOpen
  \bibfield  {author} {\bibinfo {author} {\bibfnamefont {L.~A.}\ \bibnamefont
  {Agapito}}\ and\ \bibinfo {author} {\bibfnamefont {M.}~\bibnamefont
  {Bernardi}},\ }\href {\doibase 10.1103/PhysRevB.97.235146} {\bibfield
  {journal} {\bibinfo  {journal} {Phys. Rev. B}\ }\textbf {\bibinfo {volume}
  {97}},\ \bibinfo {pages} {235146} (\bibinfo {year} {2018})}\BibitemShut
  {NoStop}%
\bibitem [{\citenamefont {Lawaetz}(1969)}]{Lawaetz}%
  \BibitemOpen
  \bibfield  {author} {\bibinfo {author} {\bibfnamefont {P.}~\bibnamefont
  {Lawaetz}},\ }\href {\doibase 10.1103/PhysRev.183.730} {\bibfield  {journal}
  {\bibinfo  {journal} {Phys. Rev.}\ }\textbf {\bibinfo {volume} {183}},\
  \bibinfo {pages} {730} (\bibinfo {year} {1969})}\BibitemShut {NoStop}%
\bibitem [{\citenamefont {Vogl}(1976)}]{Vogl1976}%
  \BibitemOpen
  \bibfield  {author} {\bibinfo {author} {\bibfnamefont {P.}~\bibnamefont
  {Vogl}},\ }\href {\doibase 10.1103/PhysRevB.13.694} {\bibfield  {journal}
  {\bibinfo  {journal} {Phys. Rev. B}\ }\textbf {\bibinfo {volume} {13}},\
  \bibinfo {pages} {694} (\bibinfo {year} {1976})}\BibitemShut {NoStop}%
\bibitem [{\citenamefont {Stengel}(2013)}]{Stengel2013}%
  \BibitemOpen
  \bibfield  {author} {\bibinfo {author} {\bibfnamefont {M.}~\bibnamefont
  {Stengel}},\ }\href {\doibase 10.1103/PhysRevB.88.174106} {\bibfield
  {journal} {\bibinfo  {journal} {Phys. Rev. B}\ }\textbf {\bibinfo {volume}
  {88}},\ \bibinfo {pages} {174106} (\bibinfo {year} {2013})}\BibitemShut
  {NoStop}%
\bibitem [{\citenamefont {Dreyer}\ \emph {et~al.}(2018)\citenamefont {Dreyer},
  \citenamefont {Stengel},\ and\ \citenamefont {Vanderbilt}}]{Dreyer2018}%
  \BibitemOpen
  \bibfield  {author} {\bibinfo {author} {\bibfnamefont {C.~E.}\ \bibnamefont
  {Dreyer}}, \bibinfo {author} {\bibfnamefont {M.}~\bibnamefont {Stengel}}, \
  and\ \bibinfo {author} {\bibfnamefont {D.}~\bibnamefont {Vanderbilt}},\
  }\href {\doibase 10.1103/PhysRevB.98.075153} {\bibfield  {journal} {\bibinfo
  {journal} {Phys. Rev. B}\ }\textbf {\bibinfo {volume} {98}},\ \bibinfo
  {pages} {075153} (\bibinfo {year} {2018})}\BibitemShut {NoStop}%
\bibitem [{\citenamefont {Royo}\ and\ \citenamefont
  {Stengel}(2019)}]{Royo2019}%
  \BibitemOpen
  \bibfield  {author} {\bibinfo {author} {\bibfnamefont {M.}~\bibnamefont
  {Royo}}\ and\ \bibinfo {author} {\bibfnamefont {M.}~\bibnamefont {Stengel}},\
  }\href {\doibase 10.1103/PhysRevX.9.021050} {\bibfield  {journal} {\bibinfo
  {journal} {Phys. Rev. X}\ }\textbf {\bibinfo {volume} {9}},\ \bibinfo {pages}
  {021050} (\bibinfo {year} {2019})}\BibitemShut {NoStop}%
\bibitem [{\citenamefont {Sjakste}\ \emph {et~al.}(2015)\citenamefont
  {Sjakste}, \citenamefont {Vast}, \citenamefont {Calandra},\ and\
  \citenamefont {Mauri}}]{sjaksteWannier2015}%
  \BibitemOpen
  \bibfield  {author} {\bibinfo {author} {\bibfnamefont {J.}~\bibnamefont
  {Sjakste}}, \bibinfo {author} {\bibfnamefont {N.}~\bibnamefont {Vast}},
  \bibinfo {author} {\bibfnamefont {M.}~\bibnamefont {Calandra}}, \ and\
  \bibinfo {author} {\bibfnamefont {F.}~\bibnamefont {Mauri}},\ }\href
  {\doibase 10.1103/PhysRevB.92.054307} {\bibfield  {journal} {\bibinfo
  {journal} {Phys. Rev. B}\ }\textbf {\bibinfo {volume} {92}},\ \bibinfo
  {pages} {054307} (\bibinfo {year} {2015})}\BibitemShut {NoStop}%
\bibitem [{\citenamefont {Verdi}\ and\ \citenamefont
  {Giustino}(2015)}]{verdiFr2015}%
  \BibitemOpen
  \bibfield  {author} {\bibinfo {author} {\bibfnamefont {C.}~\bibnamefont
  {Verdi}}\ and\ \bibinfo {author} {\bibfnamefont {F.}~\bibnamefont
  {Giustino}},\ }\href {\doibase 10.1103/PhysRevLett.115.176401} {\bibfield
  {journal} {\bibinfo  {journal} {Phys. Rev. Lett.}\ }\textbf {\bibinfo
  {volume} {115}},\ \bibinfo {pages} {176401} (\bibinfo {year}
  {2015})}\BibitemShut {NoStop}%
\bibitem [{\citenamefont {Mahan}(2011)}]{mahanCondensed2011}%
  \BibitemOpen
  \bibfield  {author} {\bibinfo {author} {\bibfnamefont {G.~D.}\ \bibnamefont
  {Mahan}},\ }\href@noop {} {\emph {\bibinfo {title} {Condensed Matter in a
  Nutshell}}}\ (\bibinfo  {publisher} {{Princeton University Press}},\ \bibinfo
  {year} {2011})\BibitemShut {NoStop}%
\bibitem [{\citenamefont {Fr{\"o}hlich}(1954)}]{frohlichElectrons1954}%
  \BibitemOpen
  \bibfield  {author} {\bibinfo {author} {\bibfnamefont {H.}~\bibnamefont
  {Fr{\"o}hlich}},\ }\href {\doibase 10.1080/00018735400101213} {\bibfield
  {journal} {\bibinfo  {journal} {Advances in Physics}\ }\textbf {\bibinfo
  {volume} {3}},\ \bibinfo {pages} {325} (\bibinfo {year} {1954})}\BibitemShut
  {NoStop}%
\bibitem [{\citenamefont {Born}\ and\ \citenamefont
  {Huang}(1954)}]{bornDynamical1954}%
  \BibitemOpen
  \bibfield  {author} {\bibinfo {author} {\bibfnamefont {M.}~\bibnamefont
  {Born}}\ and\ \bibinfo {author} {\bibfnamefont {K.}~\bibnamefont {Huang}},\
  }\href@noop {} {\emph {\bibinfo {title} {Dynamical Theory of Crystal
  Lattices}}}\ (\bibinfo  {publisher} {{Oxford University Press}},\ \bibinfo
  {address} {{Oxford}},\ \bibinfo {year} {1954})\BibitemShut {NoStop}%
\bibitem [{\citenamefont {Griffiths}(2017)}]{Griffiths}%
  \BibitemOpen
  \bibfield  {author} {\bibinfo {author} {\bibfnamefont {D.~J.}\ \bibnamefont
  {Griffiths}},\ }\href@noop {} {\emph {\bibinfo {title} {Introduction to
  Electrodynamics}}}\ (\bibinfo  {publisher} {Cambridge University Press},\
  \bibinfo {year} {2017})\BibitemShut {NoStop}%
\bibitem [{\citenamefont {Pizzi}\ \emph {et~al.}(2020)\citenamefont {Pizzi},
  \citenamefont {Vitale}, \citenamefont {Arita}, \citenamefont {Bl{\"u}gel},
  \citenamefont {Freimuth}, \citenamefont {G{\'e}ranton}, \citenamefont
  {Gibertini}, \citenamefont {Gresch}, \citenamefont {Johnson}, \citenamefont
  {Koretsune}, \citenamefont {{Iba{\~n}ez-Azpiroz}}, \citenamefont {Lee},
  \citenamefont {Lihm}, \citenamefont {Marchand}, \citenamefont {Marrazzo},
  \citenamefont {Mokrousov}, \citenamefont {Mustafa}, \citenamefont {Nohara},
  \citenamefont {Nomura}, \citenamefont {Paulatto}, \citenamefont {Ponc{\'e}},
  \citenamefont {Ponweiser}, \citenamefont {Qiao}, \citenamefont {Th{\"o}le},
  \citenamefont {Tsirkin}, \citenamefont {Wierzbowska}, \citenamefont
  {Marzari}, \citenamefont {Vanderbilt}, \citenamefont {Souza}, \citenamefont
  {Mostofi},\ and\ \citenamefont {Yates}}]{pizziWannier902020}%
  \BibitemOpen
  \bibfield  {author} {\bibinfo {author} {\bibfnamefont {G.}~\bibnamefont
  {Pizzi}}, \bibinfo {author} {\bibfnamefont {V.}~\bibnamefont {Vitale}},
  \bibinfo {author} {\bibfnamefont {R.}~\bibnamefont {Arita}}, \bibinfo
  {author} {\bibfnamefont {S.}~\bibnamefont {Bl{\"u}gel}}, \bibinfo {author}
  {\bibfnamefont {F.}~\bibnamefont {Freimuth}}, \bibinfo {author}
  {\bibfnamefont {G.}~\bibnamefont {G{\'e}ranton}}, \bibinfo {author}
  {\bibfnamefont {M.}~\bibnamefont {Gibertini}}, \bibinfo {author}
  {\bibfnamefont {D.}~\bibnamefont {Gresch}}, \bibinfo {author} {\bibfnamefont
  {C.}~\bibnamefont {Johnson}}, \bibinfo {author} {\bibfnamefont
  {T.}~\bibnamefont {Koretsune}}, \bibinfo {author} {\bibfnamefont
  {J.}~\bibnamefont {{Iba{\~n}ez-Azpiroz}}}, \bibinfo {author} {\bibfnamefont
  {H.}~\bibnamefont {Lee}}, \bibinfo {author} {\bibfnamefont {J.-M.}\
  \bibnamefont {Lihm}}, \bibinfo {author} {\bibfnamefont {D.}~\bibnamefont
  {Marchand}}, \bibinfo {author} {\bibfnamefont {A.}~\bibnamefont {Marrazzo}},
  \bibinfo {author} {\bibfnamefont {Y.}~\bibnamefont {Mokrousov}}, \bibinfo
  {author} {\bibfnamefont {J.~I.}\ \bibnamefont {Mustafa}}, \bibinfo {author}
  {\bibfnamefont {Y.}~\bibnamefont {Nohara}}, \bibinfo {author} {\bibfnamefont
  {Y.}~\bibnamefont {Nomura}}, \bibinfo {author} {\bibfnamefont
  {L.}~\bibnamefont {Paulatto}}, \bibinfo {author} {\bibfnamefont
  {S.}~\bibnamefont {Ponc{\'e}}}, \bibinfo {author} {\bibfnamefont
  {T.}~\bibnamefont {Ponweiser}}, \bibinfo {author} {\bibfnamefont
  {J.}~\bibnamefont {Qiao}}, \bibinfo {author} {\bibfnamefont {F.}~\bibnamefont
  {Th{\"o}le}}, \bibinfo {author} {\bibfnamefont {S.~S.}\ \bibnamefont
  {Tsirkin}}, \bibinfo {author} {\bibfnamefont {M.}~\bibnamefont
  {Wierzbowska}}, \bibinfo {author} {\bibfnamefont {N.}~\bibnamefont
  {Marzari}}, \bibinfo {author} {\bibfnamefont {D.}~\bibnamefont {Vanderbilt}},
  \bibinfo {author} {\bibfnamefont {I.}~\bibnamefont {Souza}}, \bibinfo
  {author} {\bibfnamefont {A.~A.}\ \bibnamefont {Mostofi}}, \ and\ \bibinfo
  {author} {\bibfnamefont {J.~R.}\ \bibnamefont {Yates}},\ }\href {\doibase
  10.1088/1361-648X/ab51ff} {\bibfield  {journal} {\bibinfo  {journal} {J.
  Phys.: Condens. Matter}\ }\textbf {\bibinfo {volume} {32}},\ \bibinfo {pages}
  {165902} (\bibinfo {year} {2020})}\BibitemShut {NoStop}%
\bibitem [{\citenamefont {Jhalani}\ \emph {et~al.}()\citenamefont {Jhalani},
  \citenamefont {Zhou}, \citenamefont {Park}, \citenamefont {Dreyer},\ and\
  \citenamefont {Bernardi}}]{Jhalani2020}%
  \BibitemOpen
  \bibfield  {author} {\bibinfo {author} {\bibfnamefont {V.~A.}\ \bibnamefont
  {Jhalani}}, \bibinfo {author} {\bibfnamefont {J.-J.}\ \bibnamefont {Zhou}},
  \bibinfo {author} {\bibfnamefont {J.}~\bibnamefont {Park}}, \bibinfo {author}
  {\bibfnamefont {C.~E.}\ \bibnamefont {Dreyer}}, \ and\ \bibinfo {author}
  {\bibfnamefont {M.}~\bibnamefont {Bernardi}},\ }\href@noop {} {\ }\Eprint
  {http://arxiv.org/abs/2002.08351} {arXiv:2002.08351} \BibitemShut {NoStop}%
\bibitem [{\citenamefont {Giannozzi}\ \emph {et~al.}(2009)\citenamefont
  {Giannozzi}, \citenamefont {Baroni}, \citenamefont {Bonini}, \citenamefont
  {Calandra}, \citenamefont {Car}, \citenamefont {Cavazzoni}, \citenamefont
  {Ceresoli}, \citenamefont {Chiarotti}, \citenamefont {Cococcioni},
  \citenamefont {Dabo}, \citenamefont {Corso}, \citenamefont {de~Gironcoli},
  \citenamefont {Fabris}, \citenamefont {Fratesi}, \citenamefont {Gebauer},
  \citenamefont {Gerstmann}, \citenamefont {Gougoussis}, \citenamefont
  {Kokalj}, \citenamefont {Lazzeri}, \citenamefont {{Martin-Samos}},
  \citenamefont {Marzari}, \citenamefont {Mauri}, \citenamefont {Mazzarello},
  \citenamefont {Paolini}, \citenamefont {Pasquarello}, \citenamefont
  {Paulatto}, \citenamefont {Sbraccia}, \citenamefont {Scandolo}, \citenamefont
  {Sclauzero}, \citenamefont {Seitsonen}, \citenamefont {Smogunov},
  \citenamefont {Umari},\ and\ \citenamefont
  {Wentzcovitch}}]{giannozziQUANTUM2009}%
  \BibitemOpen
  \bibfield  {author} {\bibinfo {author} {\bibfnamefont {P.}~\bibnamefont
  {Giannozzi}}, \bibinfo {author} {\bibfnamefont {S.}~\bibnamefont {Baroni}},
  \bibinfo {author} {\bibfnamefont {N.}~\bibnamefont {Bonini}}, \bibinfo
  {author} {\bibfnamefont {M.}~\bibnamefont {Calandra}}, \bibinfo {author}
  {\bibfnamefont {R.}~\bibnamefont {Car}}, \bibinfo {author} {\bibfnamefont
  {C.}~\bibnamefont {Cavazzoni}}, \bibinfo {author} {\bibfnamefont
  {D.}~\bibnamefont {Ceresoli}}, \bibinfo {author} {\bibfnamefont {G.~L.}\
  \bibnamefont {Chiarotti}}, \bibinfo {author} {\bibfnamefont {M.}~\bibnamefont
  {Cococcioni}}, \bibinfo {author} {\bibfnamefont {I.}~\bibnamefont {Dabo}},
  \bibinfo {author} {\bibfnamefont {A.~D.}\ \bibnamefont {Corso}}, \bibinfo
  {author} {\bibfnamefont {S.}~\bibnamefont {de~Gironcoli}}, \bibinfo {author}
  {\bibfnamefont {S.}~\bibnamefont {Fabris}}, \bibinfo {author} {\bibfnamefont
  {G.}~\bibnamefont {Fratesi}}, \bibinfo {author} {\bibfnamefont
  {R.}~\bibnamefont {Gebauer}}, \bibinfo {author} {\bibfnamefont
  {U.}~\bibnamefont {Gerstmann}}, \bibinfo {author} {\bibfnamefont
  {C.}~\bibnamefont {Gougoussis}}, \bibinfo {author} {\bibfnamefont
  {A.}~\bibnamefont {Kokalj}}, \bibinfo {author} {\bibfnamefont
  {M.}~\bibnamefont {Lazzeri}}, \bibinfo {author} {\bibfnamefont
  {L.}~\bibnamefont {{Martin-Samos}}}, \bibinfo {author} {\bibfnamefont
  {N.}~\bibnamefont {Marzari}}, \bibinfo {author} {\bibfnamefont
  {F.}~\bibnamefont {Mauri}}, \bibinfo {author} {\bibfnamefont
  {R.}~\bibnamefont {Mazzarello}}, \bibinfo {author} {\bibfnamefont
  {S.}~\bibnamefont {Paolini}}, \bibinfo {author} {\bibfnamefont
  {A.}~\bibnamefont {Pasquarello}}, \bibinfo {author} {\bibfnamefont
  {L.}~\bibnamefont {Paulatto}}, \bibinfo {author} {\bibfnamefont
  {C.}~\bibnamefont {Sbraccia}}, \bibinfo {author} {\bibfnamefont
  {S.}~\bibnamefont {Scandolo}}, \bibinfo {author} {\bibfnamefont
  {G.}~\bibnamefont {Sclauzero}}, \bibinfo {author} {\bibfnamefont {A.~P.}\
  \bibnamefont {Seitsonen}}, \bibinfo {author} {\bibfnamefont {A.}~\bibnamefont
  {Smogunov}}, \bibinfo {author} {\bibfnamefont {P.}~\bibnamefont {Umari}}, \
  and\ \bibinfo {author} {\bibfnamefont {R.~M.}\ \bibnamefont {Wentzcovitch}},\
  }\href {\doibase 10.1088/0953-8984/21/39/395502} {\bibfield  {journal}
  {\bibinfo  {journal} {J. Phys.: Condens. Matter}\ }\textbf {\bibinfo {volume}
  {21}},\ \bibinfo {pages} {395502} (\bibinfo {year} {2009})}\BibitemShut
  {NoStop}%
\bibitem [{\citenamefont {{van Setten}}\ \emph {et~al.}(2018)\citenamefont
  {{van Setten}}, \citenamefont {Giantomassi}, \citenamefont {Bousquet},
  \citenamefont {Verstraete}, \citenamefont {Hamann}, \citenamefont {Gonze},\
  and\ \citenamefont {Rignanese}}]{vansettenPseudoDojo2018}%
  \BibitemOpen
  \bibfield  {author} {\bibinfo {author} {\bibfnamefont {M.~J.}\ \bibnamefont
  {{van Setten}}}, \bibinfo {author} {\bibfnamefont {M.}~\bibnamefont
  {Giantomassi}}, \bibinfo {author} {\bibfnamefont {E.}~\bibnamefont
  {Bousquet}}, \bibinfo {author} {\bibfnamefont {M.~J.}\ \bibnamefont
  {Verstraete}}, \bibinfo {author} {\bibfnamefont {D.~R.}\ \bibnamefont
  {Hamann}}, \bibinfo {author} {\bibfnamefont {X.}~\bibnamefont {Gonze}}, \
  and\ \bibinfo {author} {\bibfnamefont {G.~M.}\ \bibnamefont {Rignanese}},\
  }\href {\doibase 10.1016/j.cpc.2018.01.012} {\bibfield  {journal} {\bibinfo
  {journal} {Comput. Phys. Commun.}\ }\textbf {\bibinfo {volume} {226}},\
  \bibinfo {pages} {39} (\bibinfo {year} {2018})}\BibitemShut {NoStop}%
\bibitem [{\citenamefont {Mahan}(2000)}]{mahanManyParticle2000}%
  \BibitemOpen
  \bibfield  {author} {\bibinfo {author} {\bibfnamefont {G.~D.}\ \bibnamefont
  {Mahan}},\ }\href {\doibase 10.1007/978-1-4757-5714-9} {\emph {\bibinfo
  {title} {Many-{{Particle Physics}}}}},\ \bibinfo {edition} {3rd}\ ed.\
  (\bibinfo  {publisher} {{Springer US}},\ \bibinfo {year} {2000})\BibitemShut
  {NoStop}%
\bibitem [{\citenamefont {Bl{\"o}chl}\ \emph {et~al.}(1994)\citenamefont
  {Bl{\"o}chl}, \citenamefont {Jepsen},\ and\ \citenamefont
  {Andersen}}]{blochlImproved1994}%
  \BibitemOpen
  \bibfield  {author} {\bibinfo {author} {\bibfnamefont {P.~E.}\ \bibnamefont
  {Bl{\"o}chl}}, \bibinfo {author} {\bibfnamefont {O.}~\bibnamefont {Jepsen}},
  \ and\ \bibinfo {author} {\bibfnamefont {O.~K.}\ \bibnamefont {Andersen}},\
  }\href {\doibase 10.1103/PhysRevB.49.16223} {\bibfield  {journal} {\bibinfo
  {journal} {Phys. Rev. B}\ }\textbf {\bibinfo {volume} {49}},\ \bibinfo
  {pages} {16223} (\bibinfo {year} {1994})}\BibitemShut {NoStop}%
\bibitem [{\citenamefont {Bernardini}\ \emph {et~al.}(1997)\citenamefont
  {Bernardini}, \citenamefont {Fiorentini},\ and\ \citenamefont
  {Vanderbilt}}]{Bernardini1997}%
  \BibitemOpen
  \bibfield  {author} {\bibinfo {author} {\bibfnamefont {F.}~\bibnamefont
  {Bernardini}}, \bibinfo {author} {\bibfnamefont {V.}~\bibnamefont
  {Fiorentini}}, \ and\ \bibinfo {author} {\bibfnamefont {D.}~\bibnamefont
  {Vanderbilt}},\ }\href {\doibase 10.1103/PhysRevB.56.R10024} {\bibfield
  {journal} {\bibinfo  {journal} {Phys. Rev. B}\ }\textbf {\bibinfo {volume}
  {56}},\ \bibinfo {pages} {R10024} (\bibinfo {year} {1997})}\BibitemShut
  {NoStop}%
\bibitem [{\citenamefont {{S{\'a}ghi-Szab{\'o}}}\ \emph
  {et~al.}(1998)\citenamefont {{S{\'a}ghi-Szab{\'o}}}, \citenamefont {Cohen},\
  and\ \citenamefont {Krakauer}}]{saghi-szaboFirstPrinciples1998}%
  \BibitemOpen
  \bibfield  {author} {\bibinfo {author} {\bibfnamefont {G.}~\bibnamefont
  {{S{\'a}ghi-Szab{\'o}}}}, \bibinfo {author} {\bibfnamefont {R.~E.}\
  \bibnamefont {Cohen}}, \ and\ \bibinfo {author} {\bibfnamefont
  {H.}~\bibnamefont {Krakauer}},\ }\href {\doibase 10.1103/PhysRevLett.80.4321}
  {\bibfield  {journal} {\bibinfo  {journal} {Phys. Rev. Lett.}\ }\textbf
  {\bibinfo {volume} {80}},\ \bibinfo {pages} {4321} (\bibinfo {year}
  {1998})}\BibitemShut {NoStop}%
\bibitem [{\citenamefont {Calandra}\ \emph {et~al.}(2010)\citenamefont
  {Calandra}, \citenamefont {Profeta},\ and\ \citenamefont
  {Mauri}}]{calandraAdiabatic2010}%
  \BibitemOpen
  \bibfield  {author} {\bibinfo {author} {\bibfnamefont {M.}~\bibnamefont
  {Calandra}}, \bibinfo {author} {\bibfnamefont {G.}~\bibnamefont {Profeta}}, \
  and\ \bibinfo {author} {\bibfnamefont {F.}~\bibnamefont {Mauri}},\ }\href
  {\doibase 10.1103/PhysRevB.82.165111} {\bibfield  {journal} {\bibinfo
  {journal} {Phys. Rev. B}\ }\textbf {\bibinfo {volume} {82}},\ \bibinfo
  {pages} {165111} (\bibinfo {year} {2010})}\BibitemShut {NoStop}%
\bibitem [{\citenamefont {Brunin}\ \emph {et~al.}()\citenamefont {Brunin},
  \citenamefont {Miranda}, \citenamefont {Giantomassi}, \citenamefont {Royo},
  \citenamefont {Stengel}, \citenamefont {Verstraete}, \citenamefont {Gonze},
  \citenamefont {Rignanese},\ and\ \citenamefont {Hautier}}]{Brunin2020}%
  \BibitemOpen
  \bibfield  {author} {\bibinfo {author} {\bibfnamefont {G.}~\bibnamefont
  {Brunin}}, \bibinfo {author} {\bibfnamefont {H.~P.~C.}\ \bibnamefont
  {Miranda}}, \bibinfo {author} {\bibfnamefont {M.}~\bibnamefont
  {Giantomassi}}, \bibinfo {author} {\bibfnamefont {M.}~\bibnamefont {Royo}},
  \bibinfo {author} {\bibfnamefont {M.}~\bibnamefont {Stengel}}, \bibinfo
  {author} {\bibfnamefont {M.~J.}\ \bibnamefont {Verstraete}}, \bibinfo
  {author} {\bibfnamefont {X.}~\bibnamefont {Gonze}}, \bibinfo {author}
  {\bibfnamefont {G.-M.}\ \bibnamefont {Rignanese}}, \ and\ \bibinfo {author}
  {\bibfnamefont {G.}~\bibnamefont {Hautier}},\ }\href@noop {} {\ }\Eprint
  {http://arxiv.org/abs/2002.00628} {arXiv:2002.00628} \BibitemShut {NoStop}%
\end{thebibliography}

\end{document}